\documentclass[aps,prd,twocolumn,showpacs,floatfix,nofootinbib,superscriptaddress]{revtex4-1}
\usepackage{amsmath}
\usepackage{amssymb}
\usepackage{graphicx}
\usepackage{hyperref}
\usepackage{color}
\usepackage{mathtools}
\usepackage{times}

\usepackage[normalem]{ulem}

\newcommand{\apjl}{Astrophys. J. Lett.}

\newcommand{\apjs}{Astrophys. J. Suppl. Sear.}
\newcommand{\aap}{Astron. Astrophys.}
\newcommand{\mnras}{Mon. Not. R. Astron. Soc.}

\newcommand{\jcap}{J. Cosmol. Astropart. Phys.}

\newcommand{\beeq}{\begin{equation}}
\newcommand{\eneq}{\end{equation}}
\newcommand{\bear}{\begin{eqnarray}}
\newcommand{\enar}{\end{eqnarray}}
\newcommand{\nnn}{\nonumber \\}

\newcommand{\RA}{\rightarrow}
\newcommand{\pa}{\partial}

\newcommand{\HH}{\mathcal{H}}   
       
\newcommand{\mpc}{{\rm Mpc}}
\newcommand{\hmpc}{{h^{-1}\mpc}}

\newcommand{\TT}{\mathcal{T}} 
\newcommand{\TTT}{\mathbb{T}}

\begin{document}

\title{Monopole Fluctuation of the CMB and its Gauge Invariance}

\author{Sandra Baumgartner}
\email{sandra.baumgartner@uzh.ch}
\affiliation{Center for Theoretical Astrophysics and Cosmology,
Institute for Computational Science,
University of Z\"urich, Winterthurerstrasse 190,
CH-8057, Z\"urich, Switzerland}
\author{Jaiyul Yoo}
\affiliation{Center for Theoretical Astrophysics and Cosmology,
Institute for Computational Science,
University of Z\"urich, Winterthurerstrasse 190,
CH-8057, Z\"urich, Switzerland}
\affiliation{Physics Institute, University of Z\"urich,
Winterthurerstrasse 190, CH-8057, Z\"urich, Switzerland}

\date{\today}

\begin{abstract}
The standard theoretical description~$\Theta(\hat n)$ 
of the observed CMB temperature anisotropies is gauge-dependent. It is, 
however, well known that the gauge mode is limited to the monopole and
the higher angular multipoles~$\Theta_l$ ($l\geq1$) are gauge-invariant.
Several attempts have been made in the past to properly define the monopole 
fluctuation, but the resulting values of the monopole power~$C_0$ 
are infinite due to the infrared divergences. The infrared divergences 
arise from the contribution of the uniform gravitational potential to
the monopole fluctuation, in violation of the
equivalence principle. Here we present the gauge-invariant theoretical
description of the observed CMB temperature anisotropies
and compute the monopole power $C_0=1.66\times10^{-9}$ in 
a $\Lambda$CDM model. While the gauge-dependence in the standard calculations 
originates from the ambiguity in defining the hypersurface for the background
CMB temperature~$\bar T$ today, it is in fact well defined and one of the
fundamental cosmological parameters.
We argue that once the cosmological parameters are chosen, the monopole fluctuation can be unambiguously inferred from the angle-average of the observed CMB temperature, making it a model-dependent ``observable''.
Adopting simple approximations for the anisotropy formation, 
we derive a gauge-invariant analytical expression for the
observed CMB temperature anisotropies to study the CMB monopole fluctuation
and the cancellation of the uniform gravitational potential contributions on large scales.
\end{abstract}

\maketitle

\section{Introduction} 
Two years after the discovery of the cosmic microwave background (CMB) radiation by Penzias and Wilson in 1965 \cite{1965ApJ...142..419P}, Sachs and Wolfe published their pioneering work \cite{Sachs_Wolfe_1967} about the formation of the CMB temperature anisotropies. 
Since then, the theoretical description of the CMB anisotropies has been 
extensively studied in many works (see, e.g., \cite{1984ApJ...285L..45B,1987MNRAS.226..655B,1994ApJ...435L..87S,1995ApJ...444..489H,1997PhRvD..56..596H,1998PhRvD..57.3290H}).
The first detection of the CMB anisotropies by the Cosmic Background Explorer (COBE) satellite was announced in 1992 \cite{1992ApJ...396L...1S}, and a variety of experiments (ground-, balloon- and space-based) have been carried out since then.
The polarization of the CMB was discovered in 2002 by the Degree Angular Scale Interferometer (DASI) telescope \cite{2002Natur.420..772K}. 
Launched in 2001, the Wilkinson Microwave Anisotropy Probe (WMAP) satellite collected data for nine years, and the final data was released in 2012 \cite{2013ApJS..208...20B}.
Its successor, the Planck satellite, was launched in 2009, and 
the final data was released in a series of papers in 2018 (see, e.g., 
\cite{Planck2018}).
Measurements on small angular scales were provided by the Atacama Cosmology Telescope (ACT) \cite{2011ApJ...739...52D} and the South Pole Telescope (SPT) \cite{2011ApJ...743...28K}.
In addition to observational data, accurate numerical computations of the CMB temperature anisotropies are provided by different versions of the Boltzmann codes (e.g., \textsc{cmbfast} \cite{1996ApJ...469..437S}, \textsc{camb} \cite{CAMB}, \textsc{class} \cite{CLASS}).

Given these recent developments, we revisit the standard theoretical
description of the CMB temperature anisotropies, in particular, focusing on 
the monopole fluctuation.
Once the Universe expands and cools enough, the CMB photons decouple 
from free electrons and propagate toward the observer, so the theoretical 
description of the observed CMB temperature naturally involves the 
physical quantities at the decoupling point, along the photon path,
and at the observer position. At the linear order in perturbations,
the gravitational redshift, the Doppler effect, and the intrinsic temperature fluctuation form the contributions to the CMB temperature anisotropies 
at the decoupling position, while the CMB temperature is further affected
during the propagation of the CMB photons through the inhomogeneous Universe, 
known as the integrated Sachs-Wolfe effect \cite{Sachs_Wolfe_1967}. 
At the observer position, the observer motion is the dominant
contribution to the CMB temperature anisotropies, but there exist other 
relativistic contributions.

The standard theoretical description of the observed CMB temperature anisotropies is the temperature fluctuation at the observer position, and it is
gauge-dependent, just like any perturbation quantities. However, since the gauge transformation of the temperature fluctuation at the observer position is isotropic, only the theoretical description of the CMB monopole anisotropy is affected, and the theoretical descriptions of 
all the other higher-order angular multipoles  are gauge-invariant.
The gauge-dependence of the standard expression originates from the ambiguity in defining the hypersurface for the background CMB temperature today:
being a function of time, the background temperature in the standard background-perturbation split of the observed photon temperature depends on the coordinate system chosen to describe the time coordinate of the observer.
However, it was shown \cite{Yoo_Mitsou_Dirian_Durrer_2019} that 
the background CMB temperature today is in fact well defined and there exists no
further gauge ambiguity.

The gauge issues associated with the theoretical description of the CMB temperature anisotropies was extensively discussed in the comprehensive work \cite{Zibin_Scott_2008} by Zibin and Scott.
An analytical expression for the CMB temperature anisotropies was derived, and the resulting monopole and the dipole transfer functions were discussed in detail.
It was stated \cite{Zibin_Scott_2008}
that the choice of the observer hypersurface (or the time coordinate of the observer) is not uniquely fixed by any physical prescription and 
this ambiguity affects the monopole of the CMB temperature anisotropy.
By specifying the observer hypersurface to be one of the uniform energy density, their expression for the observed CMB temperature anisotropy is gauge-invariant, but the resulting monopole power $C_0$ is divergent. 
Other attempts have been made in the past to derive a gauge-invariant expression, though the focus was not on the monopole fluctuation (see, e.g. \cite{1999PhRvD..59f7302H}).
All the predictions for the monopole power are divergent as well.

In this work we derive a gauge-invariant analytical expression 
for the observed CMB temperature anisotropies $\hat{\Theta}(\hat n)$
with particular emphasis on the CMB monopole anisotropy and its
gauge invariance.
We find that the resulting monopole power is devoid of any divergences, 
as the uniform gravitational potential contributions to the monopole 
fluctuation on large scales are canceled 
in accordance with the equivalence principle.
Our numerical computation shows for the first time that the (finite) monopole
power is $C_0=1.66\times10^{-9}$ in a $\Lambda$CDM universe.
We investigate the validity of our expression by comparing 
to the Boltzmann codes and compare our results to the previous work.

It is often said that the monopole fluctuation is not directly observable, as it contributes to the angle average of the observed CMB temperature together with the arbitrary background temperature.
We argue that the background CMB temperature today in eq.~\eqref{eqn:Tbar} is unambiguously defined in a given model and its cosmological parameters, and in fact it is one of the fundamental cosmological parameters \cite{Yoo_Mitsou_Dirian_Durrer_2019}.
Once a choice of the cosmological parameters is made, the monopole fluctuation \textit{can} be inferred from the angle-average of the observed CMB temperature. 
Therefore, the CMB monopole fluctuation is an observable in a sense similar to the case in galaxy clustering, where the observed galaxy number density and its power spectrum are observables, once a choice of cosmological parameters is made.

The organization of this paper is as follows: 
In Sec.~\ref{subsec:GI_background_temp}, we investigate the coordinate dependence of the background photon temperature.
Our main result, a gauge-invariant analytical expression for the observed CMB temperature anisotropies, is derived in Sec.~\ref{subsec:GI_expression_temperature}.
We clarify some issues associated with both the sky average of the observed photon temperature 
and the observability of the monopole fluctuation in Sec.~\ref{subsec:SkyAverage} and decompose our analytical expression for the temperature anisotropies in terms of observed angle in Sec.~\ref{subsec:multipoles}.
We focus on the monopole and the dipole in Sections~\ref{subsec:monopole} and~\ref{subsec:dipole} respectively and numerically compute their power.
Their large-scale limits are investigated in more detail in Sec.~\ref{subsec:large_scale_limit}.
We compare our result to previous work in Sec.~\ref{subsec:comparison} and conclude with a discussion in Sec.~\ref{sec:discussion}.

\section{Observed CMB temperature} 
\label{sec:observed_temperature}
Here we present our theoretical description of the observed CMB temperature.
In Section~\ref{subsec:GI_background_temp} we discuss the coordinate
dependence of the background CMB temperature and the issues associated with it. A gauge-invariant expression 
for the observed CMB temperature anisotropy 
is then derived in Section~\ref{subsec:GI_expression_temperature}. 
We discuss the sky average of the observed temperature
and clarify some ambiguities concerning the observability of the monopole fluctuation
in Section~\ref{subsec:SkyAverage},
and derive an expression for the multipole coefficients of the observed CMB temperature anisotropy in Section \ref{subsec:multipoles}.

\subsection{Background CMB temperature $\bar{T}$}
\label{subsec:GI_background_temp}
In perturbation analysis, it is convenient to split any quantities into a 
background and a perturbation. For CMB photons in
a thermal equilibrium, the CMB temperature $T$ 
at a given spacetime position~$x^\mu$ 
can be split as
\begin{equation}
T(x^\mu)=\bar{T}(\eta)\left[1+\Theta(x^\mu)\right] \, ,
\label{temperature}
\end{equation}
where $\eta$ is the conformal time coordinate.
The background temperature $\bar{T}(\eta)$ represents the photon temperature in a homogeneous universe, and the remaining part defines the perturbation 
or the (dimensionless) temperature fluctuation 
$\Theta:=T(x^\mu)/\bar T(\eta_o)-1$.
The homogeneity and isotropy in the background universe renders the background
temperature~$\bar T(\eta)$ only a function of time and independent of 
spatial position. This implies that the value of the background CMB 
temperature~$\bar{T}(\eta)$ at a given spacetime position depends on our choice of coordinate system, or the time coordinate~$\eta$  for the given position~$x^\mu$. 
In particular,
the background temperature $\bar T(\eta_o)$ for the observer today will depend 
on the time coordinate~$\eta_o$ of the observer, where the subscript~$o$
indicates the observer position.

Given the diffeomorphism symmetry in general relativity, we can choose
any coordinate system to describe physics, and we 
consider a general coordinate transformation 
\begin{equation}
x^\mu\mapsto \tilde{x}^\mu= x^\mu+\xi^\mu \, ,\qquad 
\xi^\mu:=(\xi,L^{,\alpha})~,
\label{eqn:GeneralCoordinateTransformation}
\end{equation}
where $\alpha,\beta,\cdots$ represent the spatial indices, while $\mu,\nu,
\cdots$ represent the spacetime indices.
Under the coordinate transformation, the observer position today is then 
described by two different coordinates $x^\mu_o:=(\eta_o,x^\alpha_o)$ 
and $\tilde x^\mu_o:=(\tilde\eta_o,\tilde x^\alpha_o)$ with the relation
\beeq
\tilde \eta_o=\eta_o+\xi_o~,\qquad \tilde x^\alpha_o=x^\alpha_o+L^{,\alpha}_o
~.
\eneq
It is now apparent that the background CMB temperature~$\bar T(\eta_o)$
at the observer position in two different coordinate systems is different by
\beeq
\bar T(\tilde \eta_o)=\bar T(\eta_o)+\bar T'(\eta_o)\xi_o~,
\eneq
where the prime denotes the derivative with respect to the conformal time
and we expanded to the linear order in perturbations.
After recombination, the CMB photons free-stream, and the observer 
at~$x^\mu_o$ today measures the black-body temperature~$T(\hat n)$ of
the CMB photons along the observed direction~$\hat n$.
The fact that the observed CMB temperature~$T(\hat n)=\bar T(\eta_o)
(1+\Theta)$ arriving at the observer position~$x^\mu_o$ should 
be independent of our coordinate choice provides the consistency relation
for the temperature fluctuation~$\Theta$ under the coordinate transformation:
\beeq
\tilde \Theta(\tilde x^\mu_o)=\Theta(x^\mu_o)+\HH_o\xi_o~,
\label{thetatransform}
\eneq
and it is indeed consistent with the gauge transformation property of~$\Theta$
at a given coordinate~$x^\mu$, where $\HH$ is the conformal Hubble parameter,
$\bar T\propto1/a$,
and we suppressed the dependence of the temperature fluctuation
on the observed direction~$\hat n$.

While the observed CMB temperature~$T(\hat n)$ along the direction~$\hat n$
is independent of coordinate system, both the background~$\bar T(\eta_o)$
and the perturbation~$\Theta$ separately
depend on our choice of coordinate system.
We remove this coordinate-dependence of the background 
temperature~$\bar T(\eta_o)$
 by introducing a fixed reference time~$\bar \eta_o$
(or $\bar t_o$), independent of our coordinate choice:
\begin{equation}
\bar \eta_o:=\int_0^\infty{dz\over H(z)}~,\qquad
\bar t_o:=\int_0^\infty{dz\over H(z)(1+z)}~,
\label{referencetime}
\end{equation}
where $H(z)=(1+z)\HH$ 
is the Hubble parameter. The coordinate-independent reference 
time~$\bar t_o$ is known as the age of
the (homogeneous) Universe, and $\bar\eta_o$ is the conformal time, 
corresponding to the proper time~$\bar t_o$. Note that the exact value 
of our reference~$\bar\eta_o$ (or its theoretical prediction)
depends on our choice of cosmological parameters, but it
is independent of our choice of coordinate system to describe the observed
universe. Furthermore, it was noted \cite{Yoo_Mitsou_Dirian_Durrer_2019,2019JCAP...12..015Y} that
 the background CMB temperature~$\bar T(\bar\eta_o)$,
not $\bar T(\eta_o)$, is really the CMB temperature today
in a homogeneous universe,
corresponding to the cosmological parameter~$\omega_\gamma$, or the 
radiation density parameter.
For later convenience we define
\begin{equation}
\bar{T}:=\bar{T}(\bar{\eta}_o) \, .
\label{eqn:Tbar}
\end{equation}

Having defined the reference time~$\bar\eta_o$ in eq.~\eqref{referencetime}, 
we express the observer position in terms of~$\bar\eta_o$ to
take advantage of its coordinate independence as
\beeq
\eta_o:=\bar{\eta}_o+\delta\eta_o \, , 
\label{lapse}
\eneq
where the coordinate (time) lapse~$\delta\eta_o$ 
represents the difference of the time coordinate~$\eta_o$
of the observer in a given coordinate system,
compared to the reference time~$\bar\eta_o$.
Mind that eq.~\eqref{lapse} is not the usual background-perturbation split
as in eq.~\eqref{temperature},
rather it is a way to describe the observer time coordinate~$\eta_o$ 
in terms of
a coordinate-independent reference $\bar{\eta}_o$. Under the coordinate
transformation in eq.~\eqref{eqn:GeneralCoordinateTransformation}, we can
derive 
\beeq
\delta\eta_o\mapsto\widetilde{\delta\eta}_o=\delta\eta_o + \xi_o \, ,
\label{lapsetransform}
\eneq
based on the coordinate independence of~$\bar\eta_o$ in eq.~\eqref{referencetime}.
A similar relation can be derived for the spatial coordinate shift with respect
to the reference position \cite{Fanizza_2018}, 
but for our purposes of the linear-order
analysis only the coordinate lapse~$\delta\eta_o$ will be used.

Using the reference time~$\bar\eta_o$, the background 
temperature~$\bar T(\eta_o)$ at the observer position today is expressed as
\begin{equation}
\bar{T}(\eta_o)=\bar{T}(\bar{\eta}_o)\left(1-\mathcal{H}_o\delta\eta_o\right)
 \, ,
\label{eqn:RelationBackgroundTemperature}
\end{equation}
and hence the observed CMB temperature along~$\hat n$ is now
\beeq
T(\hat{n})= \bar{T}(\bar{\eta}_o)\left(1-\mathcal{H}_o\delta\eta_o+\Theta_o\right) \, ,
\label{obsT}
\eneq
where $\Theta_o := \Theta(x_o^\mu)$ is the temperature fluctuation at the observer position.
The coordinate independence of~$T(\hat n)$ can be readily verified by using
the gauge
transformation properties in eqs.~\eqref{thetatransform} and~\eqref{lapsetransform}, and the combination
\beeq
\hat\Theta(\hat n):=\Theta_o(\hat n)-\HH_o\delta\eta_o~,
\label{hTheta}
\eneq
is the linear-order gauge-invariant expression for 
the observed CMB temperature anisotropies. 
The coordinate lapse~$\delta\eta_o$ is independent of the observed 
direction~$\hat n$, as it is associated with the observer motion, rather
than observations of CMB photons.
Note, however, that we have not specified the observer and the gauge-invariant
expression in eq.~\eqref{hTheta} is general.

The coordinate lapse~$\delta\eta_o$ of the observer can be derived simply
by integrating the observer four-velocity 
over the observer path \cite{Yoo_2014}.
In a homogeneous universe, all the observers are stationary, and the four-velocity
is $\bar u^\mu=(1,\vec 0)/a$, where~$a$ is the scale factor.
The integration of the time component of the
observer four-velocity yields~$\bar\eta_o$ in eq.~\eqref{referencetime}.
Due to the inhomogeneity in the universe, the observer four-velocity
$u^\mu=(1-\alpha,U^\alpha)/a$ deviates from
the stationary motion in a homogeneous universe, introducing 
the coordinate lapse $\delta\eta_o$, where $U^\alpha$ is the peculiar
velocity and $\alpha$ is the metric perturbation in the time component 
(see Appendix~\ref{app:metric} for our notation convention).
At the linear order in perturbations, the coordinate lapse is derived in 
\cite{Yoo_2014} as
\begin{equation}
\delta\eta_o=-\frac{1}{a_o}\int_0^{\bar{t}_o} dt \, \alpha \, ,
\label{eqn:DeltaEta}
\end{equation}
where the integral is in fact along the motion of the observer but equivalent
to the integral along the time coordinate at the linear order.
This expression indeed satisfies the transformation relation in eq.~\eqref{lapsetransform}.

Ignoring the vector perturbation, the spatial part of the observer four-velocity
can be expressed in terms of a scalar velocity potential $v$, i.e. 
$u_\alpha=:-av_{,\alpha}$.
If we assume that the observer follows the geodesic motion
$0=u^\nu u^\mu{}_{;\nu}$, eq.~\eqref{eqn:DeltaEta} for
the coordinate lapse can be solved to yield
\beeq
\delta\eta_o=-v_o~,
\label{geodesic}
\eneq
where the semi-colon represents the covariant derivative with respect to $g_{\mu\nu}$.
As discussed, the coordinate lapse $\delta\eta_o$ in eq.~\eqref{eqn:DeltaEta} is generic for all 
observers at the linear order, and eq.~\eqref{geodesic} is also generic for all observers on
a geodesic motion. However, it depends on spatial position and the geodesic path.

\subsection{Gauge-invariant expression for the observed CMB temperature anisotropies}
\label{subsec:GI_expression_temperature}
Equation \eqref{hTheta} is the gauge-invariant description of the observed CMB temperature anisotropies
given the gauge-dependent expression of the CMB temperature
fluctuation~$\Theta$ at the observer position,
which can be obtained by evolving the Einstein-Boltzmann equation. 
Instead, we derive a simple analytic expression for the CMB temperature 
anisotropies~$\hat\Theta(\hat n)$ in eq.~\eqref{hTheta}.
Here we assume that 1) the baryon-photon fluid is tightly coupled until 
the recombination, 2) the recombination takes place instantaneously as soon as the temperature of the baryon-photon fluid reaches $T_*$ set by atomic physics, 3) the baryon-photon fluid simultaneously decouples, and 4) no further interaction occurs for the free-streaming photons. This approximation has been adopted 
in literature to gain intuitive understanding of CMB physics (see, e.g.,
\cite{1995ApJ...444..489H,1994ApJ...435L..87S,Sachs_Wolfe_1967,Zibin_Scott_2008}). 

As the universe expands, the temperature of the baryon-photon fluid cools down, and it becomes $T_*$ at some spacetime position denoted as $x^\mu_*$, at which the CMB photons decouple from the baryons. The equilibrium temperature~$T_*$ is again split
into a background $\bar{T}$ and a perturbation $\Theta$ as
\begin{equation}
T_*:= T(x^\mu_*)=\bar{T}(\eta_*)\left[1+\Theta_*\right] \, ,
\label{eqn:decouplingTemperature}
\end{equation}
where $\Theta_*:=\Theta(x^\mu_*)$ is the temperature fluctuation at $x^\mu_*$.
 Under the tight coupling approximation, the baryon-photon fluid is fully described by the density and the velocity and is devoid of any higher-order 
moments in the photon distribution such as the anisotropic pressure. 
Note that while $T_*$ is, under our approximation, a unique number set by atomic physics, the individual components $\bar{T}(\eta_*)$ and~$\Theta_*$ depend on our choice of coordinates $x^\mu_*$ to parametrize the physical spacetime position at which the baryon-photon fluid decouples. 

Once the CMB photons decouple, they free-stream and the observer measures the CMB temperature $T(\hat n)$ in the rest-frame along the observed direction $\hat n$.
With information that the observed CMB photons originate from the initial temperature $T_*$, the observed CMB temperature $T(\hat n)$ can be used to define the observed redshift $z_\text{obs}$ of the position $x^\mu_*$, at which the
photons started to free-stream toward the observer:
\begin{equation}
1+z_\text{obs}(\hat{n}) := \frac{T_*}{T(\hat{n})} \, ,
\label{eqn:redshift_relation}
\end{equation}
as the CMB photons follow the Planck distribution and Wien's displacement law states that the wavelength at the peak of the distribution is inversely proportional to the temperature. The advantage in expressing $T(\hat n)$ in terms of 
$z_\text{obs}(\hat n)$ is that we can utilize the well-known expression for the observed redshift~$z_{\rm obs}$ and understand how the observed CMB temperature~$T(\hat n)$ is affected throughout the photon propagation, instead of solving the complicated Boltzmann equation. 
Note that no quantities in eq.~\eqref{eqn:redshift_relation} depend on our
coordinate choice.
In fact, this approach to describing the observed
CMB temperature anisotropies was first developed in \cite{Sachs_Wolfe_1967},
but the modern approach in literature focuses on the Boltzmann equation as it simplifies the calculations of the angular multipoles.

The observed redshift is 
also split into a background and a perturbation as
\begin{equation}
1+z_\text{obs}(\hat{n}):=\left(1+z_*\right)\left(1+\delta z_*\right)
\, ,
\label{zobs}
\end{equation}
where the background redshift~$z_*$ of the position $x^\mu_*$ is literally
the expression of the time coordinate~$\eta_*$:
\begin{equation}
1+z_* = \frac{a(\bar{\eta}_o)}{a(\eta_*)} = \frac{\bar{T}(\eta_*)}{\bar{T}(\bar{\eta}_o)} \, .
\label{zstar}
\end{equation}
It is convention to set $a(\bar\eta_o)\equiv1$, while it is noted that the scale factor at the observer position is
\beeq
a(\eta_o)=a(\bar\eta_o)+a'(\bar\eta_o)\delta\eta_o\neq1~.
\eneq
Consequently, the time coordinate~$z_*$ of the decoupling position
differs in two different coordinates in 
eq.~\eqref{eqn:GeneralCoordinateTransformation} as
\beeq
1+\tilde z_*=(1+z_*)(1-\HH_*\xi_*)~.
\label{zstarT}
\eneq

The perturbation $\delta z_*$ in the observed redshift
can be derived by solving the geodesic equation (see, e.g., \cite{Yoo_Grimm_Mitsou_Amara_Refregier_2018}) as
\begin{eqnarray}
\delta z_*(\hat n)&=& -H_*\chi_*+\left(\mathcal{H}\delta\eta+H\chi\right)_o-\left[v_{\chi,\alpha}n^\alpha+\alpha_\chi\right]^*_o \nonumber \\
&&-\int_0^{\bar{r}_*} d\bar{r}\left(\alpha_\chi-\varphi_\chi\right)' \, ,
 \label{eqn:definition_deltaz}
\end{eqnarray}
where $n^\alpha$ is the $\alpha$-component of 
the unit directional vector~$\hat{n}$ and~$\bar r$ is the line-of-sight
distance.
The script~$*$ indicates that the quantities are evaluated at the decoupling
point with~$x^\mu_*$.
$\alpha_\chi$ and $\varphi_\chi$ are the two gauge-invariant potentials corresponding to the Bardeen variables $\Phi_A$ and $\Phi_H$ in \cite{1980PhRvD..22.1882B}, and $\chi$ is the scalar shear of the normal observer. The velocity 
potential~$v$
is combined with the scalar shear~$\chi$ to form the gauge-invariant variable $v_\chi$ (see Appendix \ref{app:metric}).
The term $\left. \alpha_\chi\right|_o^*$ (the Sachs-Wolfe effect \cite{Sachs_Wolfe_1967}) accounts for the gravitational redshift
induced by the difference in the gravitational potential at 
departure~$x^\mu_*$ and arrival~$x^\mu_o$,
and the integral term 
is the integrated Sachs-Wolfe effect that takes into account the variation in time of the scalar metric perturbations $\alpha_\chi$ and $\varphi_\chi$. 
The individual terms in eq.~\eqref{eqn:definition_deltaz} are expressed
in terms of gauge-invariant variables (such as $\alpha_\chi$)
and the gauge-invariant combination
$(\HH\delta\eta+H\chi)_o$, except the first term~$H_*\chi_*$. Therefore, the
perturbation~$\delta z_*$ is gauge-dependent and transforms as
\beeq
\widetilde{\delta z}_*=\delta z_*+\HH_*\xi_*~.
\eneq
Note, however, that together with eq.~\eqref{zstarT}, the combination
for the observed redshift in eq.~\eqref{zobs} remains unchanged, as
the coordinate transformation in
eq.~\eqref{eqn:GeneralCoordinateTransformation} describes the same
physical spacetime point of the baryon-photon decoupling in two different
coordinates and the physical observables should be independent of our
coordinate choice.

Using eqs.~\eqref{eqn:decouplingTemperature}$-$\eqref{zstar}
and noting $\bar T:=\bar T(\bar\eta_o)$, 
the observed CMB temperature can be written as
\beeq
T(\hat n)=\bar T[1+\hat\Theta(\hat n)]~,
\label{Tobs}
\eneq
and we arrive at one of our main results, or the gauge-invariant expression 
for the CMB temperature 
anisotropies~$\hat\Theta(\hat n)$ at the linear order in perturbation:
\begin{equation}
\hat\Theta(\hat{n}):=\Theta_o(\hat{n}) - \mathcal{H}_o \delta\eta_o 
= \Theta_* - \delta z_*(\hat{n}) \, .
\label{eqn:ThetaObs}
\end{equation}
The first equation states that the observed CMB temperature anisotropies
are not described by the gauge-dependent~$\Theta_o$, but the 
gauge-invariant~$\hat\Theta_o$ that includes
the correction~$\HH_o\delta\eta_o$ due to the observer position.
The second equation provides a physical description
for the observed CMB temperature anisotropies.
Owing to the fluid approximation, the temperature 
fluctuation~$\Theta_*$ at decoupling is isotropic and does not
depend on the observed direction~$\hat n$. However, 
it is evaluated at the point of decoupling $x^\mu_*$, which is
a function of the observed direction.

\subsection{Sky average of the observed CMB temperature and the observed monopole fluctuation}
\label{subsec:SkyAverage}
The observed CMB temperature~$T(\hat n)$
can be averaged over the sky to yield the mean temperature:
\beeq
\left\langle T\right\rangle_\Omega:=\int{d^2\hat n\over4\pi}~T(\hat n)
=\bar T\left[1+\hat\Theta_0\right]~,
\label{skyavg}
\eneq
where $\hat\Theta_0$ is the angle-averaged anisotropy (or monopole fluctuation)
\beeq
\hat\Theta_0:=\int{d^2\hat n\over4\pi}~\hat\Theta(\hat n)~,
\eneq
and it should not be confused with the gauge-dependent temperature 
fluctuation~$\Theta_o(\hat n)$ at the observer position 
(mind the difference in the two subscripts $0$ and $o$). In comparison,
the ensemble average of the observed CMB temperature is
\beeq
\left\langle T(\hat n)\right\rangle=\bar T~,\qquad \qquad
\left\langle \hat\Theta(\hat n)\right\rangle=0~.
\label{ensemble}
\eneq
We want to emphasize that~$\bar T$, not $\langle T\rangle_\Omega$
(or a coordinate-dependent $\bar T(\eta_o)$), is the CMB temperature today 
in a homogeneous universe, corresponding to a cosmological parameter, while
$\langle T\rangle_\Omega$ is the observed CMB temperature today upon angle
average.

Equations~\eqref{skyavg} and~\eqref{ensemble}
make it clear that the observed mean temperature~$\langle T\rangle_\Omega$,
for instance, from the COBE Far Infrared Absolute Spectrometer (FIRAS) \cite{1996_Fixsen_etal} differs from the background CMB 
temperature~$\bar T$, or the ensemble average, 
as it includes the monopole fluctuation~$\hat\Theta_0$
at our position. It was pointed
out \cite{2020_Mitsou_etal,Yoo_Mitsou_Dirian_Durrer_2019} that the
ensemble average is equivalent to the Euclidean average, including not only
the angle average over the sky, but also the spatial average over different
observer positions. Consequently, 
the mean temperature~$\langle T\rangle_\Omega$ today (or the angle average)
depends on the spatial
position of the observation due to the monopole fluctuation,
and its value alone cannot determine the cosmological 
parameter~$\omega_\gamma$ (or~$\bar T$). 
This implies that if one takes $\langle T\rangle_\Omega$ as the ``ensemble
average,'' there is {\it no} observed monopole fluctuation by construction.
In practice,
given the rms fluctuation amplitude~$\sim10^{-5}$ of the monopole computed in 
Section~\ref{subsec:monopole}, the difference between~$\langle T\rangle_\Omega$
and~$\bar T$ is negligible.

In addition, eq.~\eqref{skyavg} makes it clear that once the cosmological parameter~$\omega_\gamma$ (or~$\bar T$) is chosen,
the observed mean temperature~$\langle T\rangle_\Omega$ directly translates
into the ``observed'' monopole fluctuation~$\hat\Theta_0$. 
Although model-dependent, the monopole fluctuation \textit{can} be inferred once a cosmological model is chosen.
The resulting value is model-dependent, but independent of our choice of coordinate system. 
On the other hand, the observed photon temperature can be split according to eq.~\eqref{temperature} into the two gauge-dependent quantities $\bar{T}(\eta_o)$ and $\Theta(x^\mu_o)$, and the angle-averaged CMB temperature takes the form
\beeq
\left\langle T\right\rangle_\Omega:=\int{d^2\hat n\over4\pi}~T(\hat n)
=\bar T(\eta_o)\left[1+\Theta_0(x^\mu_o)\right]~,
\eneq
where $\Theta_0(x^\mu_o)$ is the angle average (or monopole) of the temperature
fluctuation~$\Theta(x^\mu_o)$ at the observer position. 
The background temperature~$\bar{T}(\eta_o)$ is ambiguous, as it depends on the time coordinate of the observer and therefore on the coordinate system chosen. 
Accordingly, $\Theta_0(x^\mu_o)$ is ambiguous as well.
This ambiguity in the background-perturbation split is the reason why the monopole fluctuation is often referred to be unobservable. 
However, the CMB temperature today in a homogeneous universe is correctly described by $\bar{T}(\bar{\eta}_o)$ and there is no ambiguity in its definition in a given model and its cosmological parameters.

Any measurements of cosmological observables in practice involve measurement uncertainties, and hence the cosmological parameters in a given model have uncertainties in their best-fit values, which result in uncertainties in the theoretical predictions. 
However, this aspect is rather independent from the goal of this work. 
These uncertainties in the predictions for the cosmological observables are solely due to the uncertainties in our estimates of the cosmological parameters, not due to the ambiguities in the theoretical predictions. 
The primary goal in this current investigation is to have a unique prediction for the monopole power $C_0$, given a model and its assumed cosmological parameters.

\subsection{Multipole expansion}
\label{subsec:multipoles}
The observed CMB temperature anisotropy~$\hat\Theta(\hat n)$ is traditionally
decomposed in terms of spherical harmonics~$Y_{lm}(\hat n)$ as
\beeq
\hat\Theta(\hat n)=\sum_{lm}\hat a_{lm}Y_{lm}(\hat n)~,
\eneq
and the multipole coefficients are
\beeq
\hat a_{lm}=\int d^2{\hat n}~Y_{lm}^*(\hat n)\hat\Theta(\hat n)~.
\label{alm}
\eneq
Defining the standard multipole coefficients $a_{lm}$ (without hat) in the same way 
for the gauge-dependent temperature
fluctuation~$\Theta_o(\hat n)$ at the observer position, we derive the
relation between the two different multipole coefficients
\beeq
\hat a_{lm} =a_{lm}
-\sqrt{4\pi}\mathcal{H}_o\delta\eta_o~\delta_{l0}\delta_{m0}~.
\label{eqn:a_lm_relation}
\eneq
With no angular dependence for~$\delta\eta_o$, the
difference resides only at the monopole with $l=0$, 
re-affirming that all the multipole
coefficients~$a_{lm}$ derived in literature are gauge-invariant for $l\geq1$.
The gauge-dependence of the (standard) monopole~$a_{00}$ is well-known
and also evident in eq.~\eqref{thetatransform}. However, we emphasize that
the correct monopole coefficient~$\hat a_{00}$ is indeed gauge-invariant and
well-defined.

While the difference is limited to the monopole coefficient in theory,
{\it all} the multipole coefficients are indeed affected in reality, though 
the impact is rather negligible in practice due to the small rms fluctuation
amplitude of the monopole. The standard Boltzmann codes such as \textsc{camb}~\cite{CAMB} and
\textsc{class}~\cite{CLASS} provide the multipole coefficients~$a_{lm}$ and their angular
power spectra~$C_l:=\langle|a_{lm}|^2\rangle$ with $l\geq2$, and the 
comparison to observations determines the best-fit cosmological parameters.
However, in the standard data analysis of CMB measurements, the background
CMB temperature is set equal to the observed mean temperature
$\bar T\equiv\langle T\rangle_\Omega$ by hand, and this
formally incorrect procedure results in two problems \cite{Yoo_Mitsou_Dirian_Durrer_2019}: 1)~The 
background dynamics
in our model predictions differs from the background evolution in our Universe,
unless the monopole~$\hat\Theta_0$ at our position happens to be zero by accident.
2)~By using the observed mean~$\langle T\rangle_\Omega$ as the background
temperature~$\bar T$, the angular multipole coefficients obtained from
the observations correspond to
\beeq
\hat a_{lm}^{\rm obs}={\hat a_{lm}\over1+\hat\Theta_0}~.
\eneq
While these two issues are shown \cite{Yoo_Mitsou_Dirian_Durrer_2019}
to have negligible impact on our current 
cosmological parameter analysis due to the small rms fluctuation of the 
monopole, these systematic errors  in the standard data analysis are always
present, which may become a significant component in the systematic errors
in future surveys (see, e.g., \cite{WENETAL2020} for recent discussion).

Combining eqs.~\eqref{eqn:definition_deltaz} and~\eqref{eqn:ThetaObs},
we expressed the observed CMB temperature anisotropies as
\beeq
\hat\Theta=\Theta_{\chi*}
+\left[{v_\chi}_{,\alpha}n^\alpha+\alpha_\chi\right]^*_o +H_o v_{\chi o}
+\int_0^{\bar{r}_*} \!\!\!
d\bar{r}\left(\alpha_\chi-\varphi_\chi\right)'~,
	\label{eqn:analytical_expression}
\eneq
where the temperature fluctuation~$\Theta_*$ at decoupling in 
eq.~\eqref{eqn:ThetaObs} is combined with the gauge-dependent term $H_*\chi_*$
in eq.~\eqref{eqn:definition_deltaz} to form a gauge-invariant 
temperature fluctuation~$\Theta_\chi$ in the conformal
Newtonian gauge
\beeq
\Theta_\chi:=\Theta+H\chi~,
\eneq
and we assumed that the observer motion is geodesic in eq.~\eqref{geodesic}
to form a gauge-invariant variable for the scalar velocity potential~$v_\chi$
at the observer position. Note that the decoupling position~$x^\mu_*$
is a function of the observed direction~$\hat n$.
To derive the expressions for the multipole coefficients~$\hat a_{lm}$, we
first define the transfer functions~$\TTT(\eta,k)$ for the 
gauge-invariant variables in eq.~\eqref{eqn:analytical_expression}
in terms of the primordial 
fluctuation~$\zeta(\mathbf{k})$
set at the initial condition. For instance, the transfer function 
for~$\alpha_\chi$ is then
\beeq
\alpha_\chi(\eta, \mathbf{k})=: \TTT_{\alpha_\chi}(\eta,k)
\zeta(\mathbf{k}) \, ,
	\label{eqn:decomposition_alpha1}
\eneq
providing the relation of $\alpha_\chi(\eta,\mathbf{k})$ in Fourier space
at any conformal 
time~$\eta$ to the initial condition $\zeta(\mathbf{k})$ of 
the comoving gauge curvature $\zeta:=\varphi-\mathcal{H}v$, where the initial
power spectrum is set as $\Delta_\zeta^2:=k^3P_\zeta/2\pi^2=A_s (k/k_\circ)^{n_s-1}$ in terms of the primordial fluctuation amplitude $A_s$ at pivot scale 
$k_\circ$ and the spectral index $n_s$ (see, e.g., \cite{Planck2018}).
It is well-known that on large scales $k\RA0$ the comoving gauge
curvature is conserved in time,
 and the gauge-invariant variable is then related as
\beeq
\alpha_\chi(\eta_{\rm mde})=-\frac35\zeta~,
\eneq
where we suppressed the scale dependence and considered the conformal time
in the matter-dominated era (mde). 
This implies that the transfer function has the limit
\beeq
\lim_{k\RA0}\TTT_{\alpha_\chi}(\eta_{\rm mde},k)=-\frac35~.
\eneq

Using the plane-wave expansion
\begin{equation}
e^{i\mathbf{k}\cdot\mathbf{x}} = 4\pi \sum_{l,m} i^l j_l\left(kr\right) Y_{lm}(\hat{\mathbf{k}}) Y_{lm}^*\left(\hat{\mathbf{x}}\right)
 \, ,
	\label{eqn:identity_exponential}
\end{equation}
the multipole coefficients $\hat a_{lm}$ of 
the CMB temperature anisotropies can be derived according to eq.~\eqref{alm}
as
\bear
	\label{eqn:a_lm}
\hat a_{lm}&=&4\pi i^l\int \frac{dk\, k^2}{2\pi^2}  \bigg\{
\left( \TTT_{\Theta_\chi} 
+\TTT_{\alpha_\chi}\right)_*j_l(k\bar{r}_*) \\
&&
+k \TTT_{v_\chi *} ~j_l'(k\bar{r}_*)  
 -\left(\TTT_{\alpha_\chi}-H\TTT_{v_\chi}\right)_o\delta_{l0} 
 -\frac{k}{3} \TTT_{v_\chi o}\delta_{l1}  \nnn
&&
+\int_0^{\bar{r}_*} d\bar{r}\,\left(\TTT'_{\alpha_\chi}
 -\TTT'_{\varphi_\chi}\right) j_l(k\bar r) 
\bigg\} \int \frac{d\Omega_k}{4\pi} Y_{lm}^*(\hat k) \zeta(\mathbf{k})  \, , \nonumber
\enar
where $j_l(x)$ are the spherical Bessel functions,
$\delta_{ll'}$ is the Kronecker delta, and we suppressed the $k$-dependence
of the transfer functions, while the time-dependence is indicated in terms
of subscripts. In the matter dominated Universe, the gravitational potential
is constant, and the integrated Sachs-Wolfe contribution vanishes. So,
the dominant contributions to the CMB temperature anisotropies today at
higher angular multipoles ($l\geq2$) are the temperature fluctuation with 
the gravitational potential contribution at the source plus the Doppler
effect (see, e.g., \cite{1995ApJ...444..489H,1994ApJ...435L..87S,2003moco.book.....D})
\beeq
\hat a_{lm}\propto \left(\TTT_{\Theta_\chi} 
+\TTT_{\alpha_\chi}\right)_*j_l(k\bar{r}_*) 
+k \TTT_{v_\chi *} ~j_l'(k\bar{r}_*)  ~.
\eneq

\section{Monopole and Dipole of the CMB Temperature Anisotropies}
\label{sec:MonopoleDipole}
The multipole coefficients~$\hat a_{lm}$ in observations can be summed over~$m$
to yield an estimate of the angular power spectrum~$C_l$. Its theoretical
prediction can be obtained by taking the ensemble average of the multipole
coefficients as
\beeq
C_l=\langle \left|\hat a_{lm}\right|^2\rangle = 4\pi\int d \ln k \,  
\Delta^2_\zeta\left(k\right) \left| \TT_l(k) \right|^2 
 \, ,
  \label{eqn:definition_Cl}
\eneq
where the stochasticity in~$\zeta(\mathbf{k})$ is averaged out to yield
the power spectrum by using
\beeq
\langle\zeta(\mathbf{k}_1)
\zeta(\mathbf{k}_2)\rangle={(2\pi)^3\over k^2_1}P_\zeta(k_1)\delta^D(k_1-k_2)\delta^D(\Omega_{k_1}-\Omega_{k_2})~,
\eneq
and the various effects in the curly bracket in eq.~\eqref{eqn:a_lm}
are lumped into the transfer function~$\TT_l(k)$ for the angular power spectrum
\bear
&&
\TT_l(k):=\left(\TTT_{\Theta_\chi} 
+\TTT_{\alpha_\chi}\right)_* j_l(k\bar{r}_*) 
+k \TTT_{v_\chi *} ~j_l'(k\bar{r}_*)  
 -\frac{k}{3} \TTT_{v_\chi o}\delta_{l1} 
\nnn
&&\quad
 -\left(\TTT_{\alpha_\chi}-H\TTT_{v_\chi}\right)_o\delta_{l0} 
+\int_0^{\bar{r}_*} d\bar{r}\,\left(\TTT'_{\alpha_\chi}
 -\TTT'_{\varphi_\chi}\right) j_l(k\bar r)~. ~~~~~~~~
 \label{eqn:transfer_function}
\enar
Since the primordial power spectrum~$\Delta_\zeta^2$ is nearly scale-invariant,
the shape of the transfer function~$\TT_l(k)$ contains all the information 
about the angular power spectrum~$C_l$.
Our main focus in this Section is on the monopole and the dipole transfer functions, which 
can be readily obtained from eq.~\eqref{eqn:transfer_function} with $l=0$
and~$l=1$. 

Our analytical expression describes the simple situation in which the
CMB photons in thermal equilibrium with~$T_*$ are emitted at the decoupling 
point~$x^\mu_*$ and these photons are measured by the observer at~$x^\mu_o$.
Though both the temperature and the radiation energy density are observer-dependent
quantities, the dependence arises at the second order from the Lorentz boost,
and no further specification of the source and the observer frames is 
necessary. However, the observed redshift~$z_{\rm obs}$ 
and its perturbation~$\delta z$ are affected by the motion of the source and
the observer via the linear-order 
Doppler effect. At the decoupling point~$x^\mu_*$, the source 
is described as the baryon-photon fluid, so the velocity at~$x^\mu_*$ is
the velocity of the baryon-photon fluid. For the observer motion, 
we assume that the observer is moving
together with matter (baryons and dark matter), $v_o\equiv v_m(\eta_o,k)$.
It is important to note that the velocity terms in 
eq.~ \eqref{eqn:definition_deltaz} are those specifying the rest frames at
the emission (or decoupling) and the observation of the photons; they
do not have to be the velocity potential of the same fluid at both emission
and observer points.

In Sections \ref{subsec:monopole} and \ref{subsec:dipole} we present 
our numerical computation of the monopole transfer function~$\TT_0(k)$ 
and the dipole transfer function~$\TT_1(k)$.
Section \ref{subsec:large_scale_limit} takes a closer look at 
the large-scale limit of the two transfer functions~$\TT_0(k)$ and~$\TT_1(k)$.
We then compare our results to previous work
on~$\TT_0(k)$ and~$\TT_1(k)$ in Section \ref{subsec:comparison}.
We choose the conformal Newtonian gauge ($\chi\equiv0$) 
for the numerical calculations
of our gauge-invariant expression in eq.~\eqref{eqn:transfer_function}.

To calculate the transfer functions of the individual scalar perturbation variables, we use the Cosmic Linear Anisotropy Solving System (\textsc{class}) \cite{CLASS} and the Code for Anisotropies in the Microwave Background (\textsc{camb}) \cite{CAMB}. 
We find that the difference between two Boltzmann code solvers is negligible in our calculations.
Here we adopt the $\Lambda$CDM model with the
cosmological parameters consistent with the {\it Planck}~2018 results
\cite{Planck2018}: dimensionless Hubble parameter $h=0.6732$,
the baryon density parameter $\Omega_\text{\tiny{b}} h^2=0.02299$,
the (cold) dark matter density parameter $\Omega_\text{\tiny{cdm}} h^2=0.12011$,
the reionization optical depth $\tau = 0.0543$, 
the scalar spectral index $n_s=0.96605$, and
the primordial amplitude $\ln(10^{10}A_s)=3.0448$ at
$k_\circ=0.05 \,\text{Mpc}^{-1}$.

\subsection{Monopole}
\label{subsec:monopole}
The transfer function of the monopole in the conformal Newtonian gauge is
\bear
\TT_0(k)&=&\left[\TTT_{\Theta}(\eta_*,k)
+\TTT_{\psi}(\eta_*,k)\right]j_0(k\bar{r}_*) \nnn
&&
+k \TTT_{v_\gamma}(\eta_*,k) ~j_0'(k\bar{r}_*) 
 -\TTT_{\psi}(\eta_o,k)+H_o\TTT_{v_m}(\eta_o,k) \nnn
&&
+\int_0^{\bar{r}_*} d\bar{r}\,\left[\TTT'_{\psi}(\eta,k)
 -\TTT'_{\phi}(\eta,k)\right] j_0(k\bar r)~,
	\label{eqn:transfer_function_monopole}
\enar
where $\psi:=\alpha$ and $\phi:=\varphi$ in the conformal Newtonian gauge.
Though $\psi\approx-\phi$ already at the decoupling, we use the exact transfer
functions for~$\psi$ and~$\phi$ to compute the monopole transfer 
function~$\TT_0(k)$. The monopole fluctuation is composed of the
photon temperature fluctuation~$\Theta_{*}$, 
the gravitational redshift~$\psi_*$,
and the baryon-photon velocity~$v_{\gamma*}$ at the decoupling point, 
the gravitational redshift~$\psi_o$ and the observer 
velocity potential~$v_m$ at the observer position, 
and finally the integrated Sachs-Wolfe effect (ISW).
Note that the velocity potential~$v_m$ at the observer position arises from the
coordinate lapse~$\delta\eta_o$ in eq.~\eqref{geodesic},
while the Doppler effect by the observer velocity
is absent in the monopole transfer function.
For later convenience, we refer to those contributions at the 
decoupling~$x^\mu_*$ as the source terms, while those at the observer position
as the observer terms. Although they are not individually observables, 
the decomposition into the individual components helps understand the
monopole transfer function intuitively.

\begin{figure}
\includegraphics[width=\linewidth]{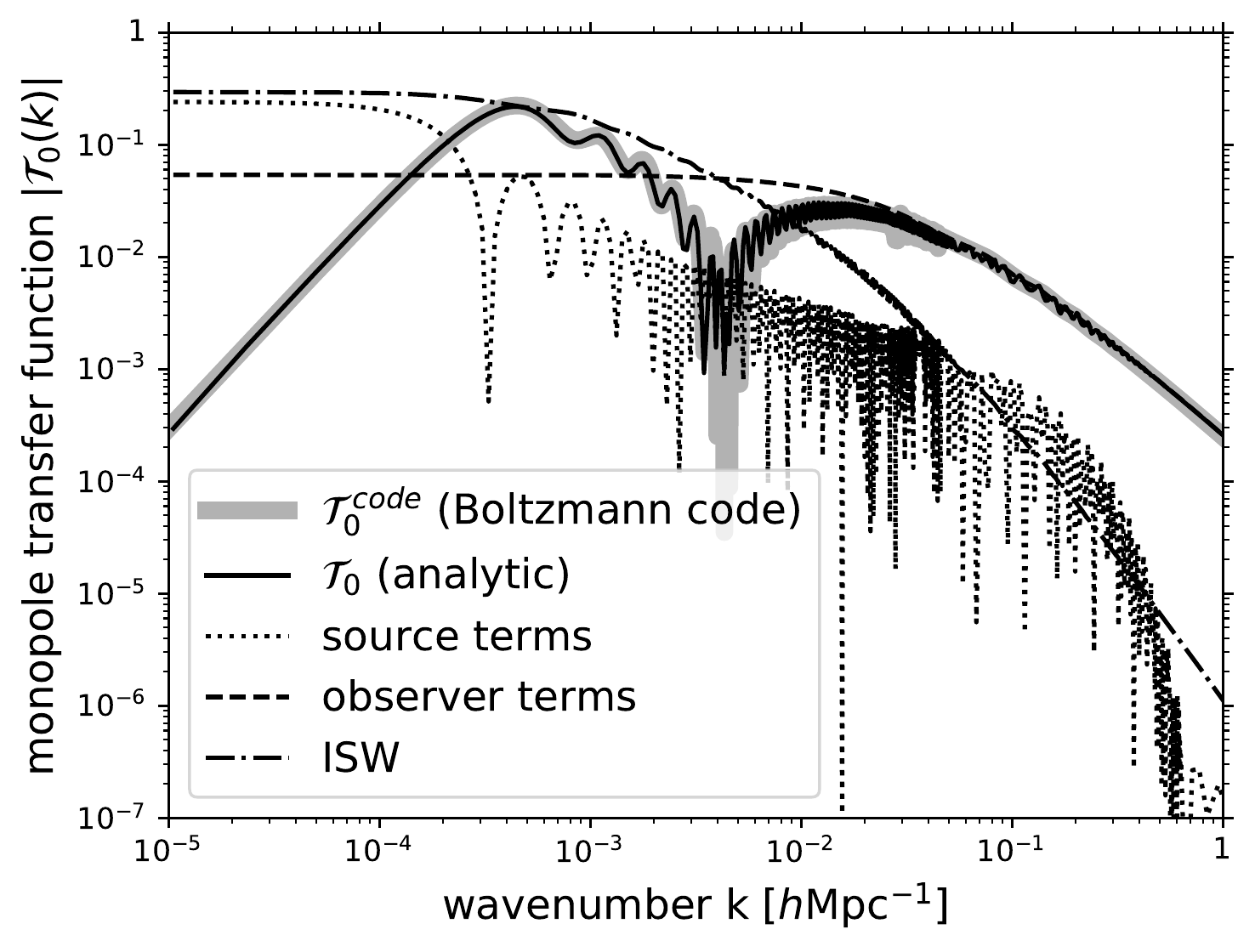}
\caption{Monopole transfer function $\TT_0$ and individual contributions as a function of the wavenumber $k$. The analytic expression (\ref{eqn:transfer_function_monopole}) derived in this paper is represented by the black solid line, while the contributions from terms evaluated at decoupling (denoted as source terms) and at the observer position are indicated by the dotted and the dashed line respectively. The dot-dashed line corresponds to the ISW contribution. The thick gray line shows the transfer function from the Boltzmann codes serving as a reference for our analytical expression.}
\label{fig:monopole}
\end{figure}

\begin{figure}
\includegraphics[width=\linewidth]{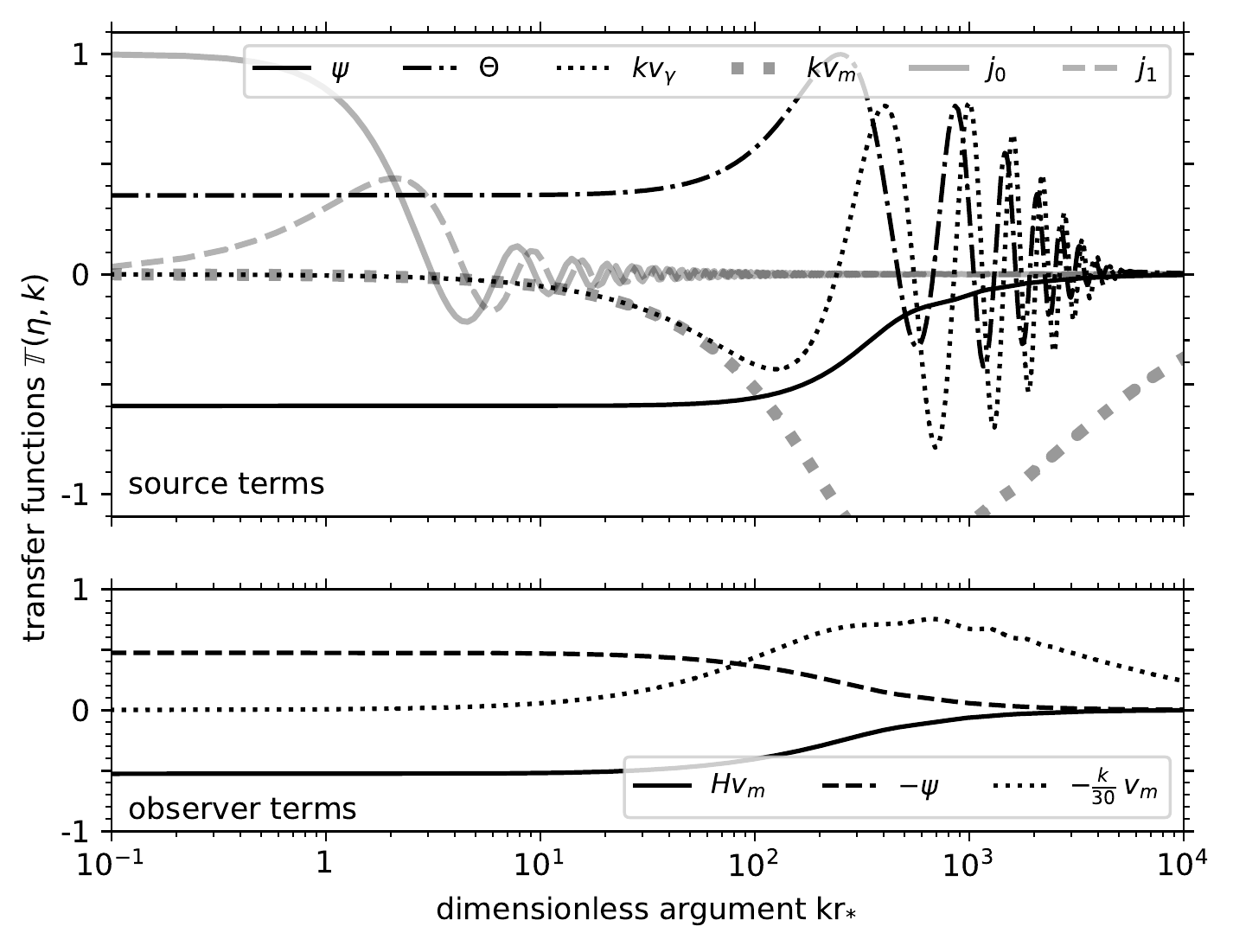}
\caption{Individual contributions to the monopole and dipole transfer functions as a function of dimensionless argument $kr_*$, where $r_*$ is the comoving distance to the last scattering surface. The range of the $x$-axis is approximately the same as in Figs. \ref{fig:monopole} and \ref{fig:dipole}. The top panel shows terms evaluated at decoupling (i.e. source terms), in addition to the two spherical Bessel functions $j_0(x)$ and $j_1(x)$. The bottom panel shows terms evaluated at the observer position. Note that the dotted curve shows $kv_m/30$ instead of $kv_m/3$, with another factor ten to fit in the plot.}
\label{fig:MonopoleContributions}
\end{figure}

Figure~\ref{fig:monopole} describes the monopole transfer function~$\TT_0(k)$
and its
individual contributions. To facilitate the comparison, we plot the absolute
values of the individual transfer functions. The integrated Sachs-Wolfe
effect (ISW; dot-dashed) and the contributions at the 
decoupling (source terms; dotted) 
are the dominant contribution to the monopole transfer function on 
large scales, while the contributions at the observer position (observer
terms; dashed) are dominant on small scales, as the source terms are
suppressed due to the spherical Bessel function.
The source terms oscillate rapidly largely due to the baryon-photon 
velocity~$v_\gamma$, and because of the Silk damping \cite{1968ApJ...151..459S} the 
baryon-photon fluctuations decay fast on small scales. On large scales,
all three contributions (source terms, observer terms, and ISW) 
are constant, and these individual contributions on super-horizon scales
($k\ll H_0=3.3\times10^{-4}h$Mpc$^{-1}$) result in infrared divergences
for the monopole power in eq.~\eqref{eqn:definition_Cl}, respectively.
However, as evident in Fig. \ref{fig:monopole}, the sum of all the individual
contributions to~$\TT_0$ (solid) falls as $k^2$ on large scales due to the cancellation
of the individually diverging contributions, and the monopole power~$C_0$
is finite and devoid of any divergences. To put it differently, the 
theoretical prediction for the observed CMB temperature today upon angle
average in eq.~\eqref{skyavg} is finite and independent of the fluctuations on 
very large scales beyond our horizon (see also \cite{Zibin_Scott_2008}).

To better understand the subtle cancellation on large scales, we show 
the individual contributions of the source terms and the observer terms in 
Fig.~\ref{fig:MonopoleContributions} in linear scale as a function
of dimensionless argument~$kr_*$ in the Bessel function, where the
distance to the decoupling is $r_*=13.87$~Gpc (or $9.34$~$h^{-1}$Gpc) at $z_*=1088$
with our fiducial cosmological parameters, so that the $x$-range in
Fig.~\ref{fig:MonopoleContributions} is approximately equivalent to
the range of the wavenumber in Fig.~\ref{fig:monopole}.
The upper panel shows
the contributions at the decoupling or the source terms that are the
photon temperature fluctuation~$\Theta$ (dot-dashed), 
the gravitational redshift~$\psi$ (solid)
and the Doppler effect~$v_\gamma$ (dotted).
Since these quantities contribute to the observed monopole fluctuation
via photon propagation from the decoupling point toward 
the observer position, all three terms are
multiplied by the spherical Bessel function $j_0(x)$ (gray solid) or its 
derivative~$j_0'(x)$, and hence suppressed by $1/x$ (or $1/x^2$ for the
Doppler term) on small scales ($x\gg10$). On large scales ($j_0\simeq1$),
the first two contributions~$\Theta$ and~$\psi$ are dominant and 
in fact constant in~$k$ but with opposite signs, giving rise to cancellation 
between the two contributions. The Doppler effect~$v_\gamma$ is again 
negligible on large scales, as it is a gradient ($\propto k$) 
and $j_0'\simeq0$.

The bottom panel shows the contributions at the observer position or
the observer terms that are made of the gravitational redshift~$\psi$ (dashed)
and the coordinate lapse~$Hv_m$ (solid). Since the observer terms are without
the spherical Bessel function, their contributions are relatively larger 
than the source terms on small scales, though the transfer functions still decay 
on small scales. The two contributions on large scales are also constant but
with opposite signs. Their amplitude is almost identical,
resulting in near cancellation, so the contribution of the observer terms
is smaller than the source terms on large scales, 
apparent in Fig.~\ref{fig:monopole}. These two contributions (the source
and the observer terms) add up to nearly cancel the positive contribution 
from the line-of-sight integration or the ISW term. 
Mind that while the gravitational potentials ($\psi$ and $\phi$) decay in time,
the line-of-sight direction increases backward in time to yield the positive contribution
of the ISW term.

To check the validity of our analytical expression,
we plot the monopole transfer function~$\TT_0^{\rm code}(k)$ 
(gray solid) in Fig.~\ref{fig:monopole} from the Boltzmann codes \textsc{class} and~\textsc{camb}. 
Numerical computation in these Boltzmann codes is performed in the synchronous
gauge, where $\alpha=\beta=0$. Using the residual gauge freedom in the
synchronous gauge (see, e.g., \cite{Yoo_2014}), an extra gauge condition
is imposed in the Boltzmann codes, 
in which $v_m=0$. So, the resulting gauge condition is indeed (dark-matter)
comoving-synchronous gauge.
The observed CMB temperature anisotropies in this gauge condition are then
\beeq
\hat\Theta(\hat n)=\Theta_o^{\rm sync}(\hat n)~,
\label{eqn:Theta_SynchronousGauge}
\eneq
from eq.~\eqref{eqn:ThetaObs}, where the coordinate lapse is vanishing
with $v_m=0$. Therefore, we can use the transfer function for 
the photon temperature fluctuation today from the Boltzmann codes
for the monopole transfer function (gray solid) in 
Fig.~\ref{fig:monopole}:
\beeq
\TT_0^{\rm code}(k)=\TTT^{\rm sync}_{\Theta}(\eta_0,k)~,
\eneq
(see Appendix \ref{app:SynchronousGauge} for more details).
Note that the transfer function~$\TTT^{\rm sync}_{\Theta}$ 
in the Boltzmann codes (hence $\TT_0^{\rm code}$)
is obtained by numerically solving the full Boltzmann equation to the present 
day without any approximations we adopted for our analytical calculations,
while our monopole transfer function~$\TT_0$ (solid curve)
is obtained by using the
analytic expression in eq.~\eqref{eqn:transfer_function_monopole}.

Figure \ref{fig:monopole} shows an astonishing agreement for the monopole transfer function (solid and gray curves) on all scales, which strongly supports that our approximations for the analytical expression capture the essential physics of the CMB anisotropy formation, at least for the monopole. Our approximations neglect collisions against free electrons after the recombination and assume a sharp transition from tight coupling to complete decoupling, none of which matter on large scales. Though we do expect the breakdown of our approximations on small scales, the monopole transfer function on small scales is dominated by the contribution at the observer position, independent of the validity of our approximations.

Using the monopole transfer function, we numerically compute the monopole power
in eq.~\eqref{eqn:definition_Cl}
\beeq
C_0=\left\langle|\hat a_{00}|^2\right\rangle=1.66\times10^{-9}~,
\eneq
and the rms fluctuation of the angle-averaged anisotropy
\beeq
\sqrt{\left\langle\hat\Theta_0^2\right\rangle}=\sqrt{{C_0\over4\pi}}
=1.15\times10^{-5}~.
\eneq
We emphasize that this is the \textit{first time} to correctly compute these values (see Sec.~\ref{subsec:comparison}).
It turns out that the rms fluctuation in our $\Lambda$CDM model is very small,
and in particular a lot smaller than the measurement uncertainty in 
the COBE FIRAS observation \cite{1996_Fixsen_etal,2009ApJ...707..916F},
\beeq
\langle T\rangle_\Omega=2.7255\,\,\pm\,\,5.7\times10^{-4}~{\rm K}~.
\eneq
This explains why the systematic errors in the standard cosmological parameter estimation are in practice
negligible despite the formally 
incorrect assumption of setting $\bar T\equiv\langle T\rangle_\Omega$
(see \cite{Yoo_Mitsou_Dirian_Durrer_2019}).

We want to emphasize again that the monopole fluctuation~$\hat\Theta_0$ \textit{is} directly ``observable'' once a choice of cosmological parameters is made, as the latter uniquely fixes the value of the background temperature~$\bar{T}$. The monopole fluctuation can then be inferred from the observed mean temperature~$\langle T\rangle_\Omega$ via eq.~\eqref{skyavg}, although the resulting value is of course model-dependent. This is discussed in more detail in Sec.~\ref{subsec:SkyAverage}.

In the past analytical calculations have been performed
in particular with the choice of the conformal Newtonian gauge.
However, the gauge-invariance of the expression was not verified,
and the coordinate lapse~$\delta\eta_o$ at the observer position is
neglected. To be fair, the focus of previous analytic calculation (see, e.g. \cite{1995ApJ...444..489H,1994ApJ...435L..87S}) is on the higher angular multipoles rather than the monopole and the dipole.
Apparent from Figs.~\ref{fig:monopole} 
and~\ref{fig:MonopoleContributions}, the monopole transfer function would
be constant in~$k$ on large scales, if~$\delta\eta_o$ is neglected,
and the resulting monopole power~$C_0$ would be infinite.
Comparison to previous work will be presented in Section \ref{subsec:comparison}.

Figure \ref{fig:MonopoleContributions} also shows the velocity potential $v_m$ (gray dotted) of the dark matter at the decoupling point. 
On large scales, it is identical to the photon velocity $v_\gamma$ (and hence the baryon velocity $v_b=v_\gamma$), but it deviates
significantly on small scales, as dark matter decoupled in the early universe and evolved separately from the baryon-photon plasma.
However, the impact of using $v_m$ instead of $v_\gamma$ on the monopole transfer function $\TT_0(k)$ is negligible.

\begin{figure}
\includegraphics[width=\linewidth]{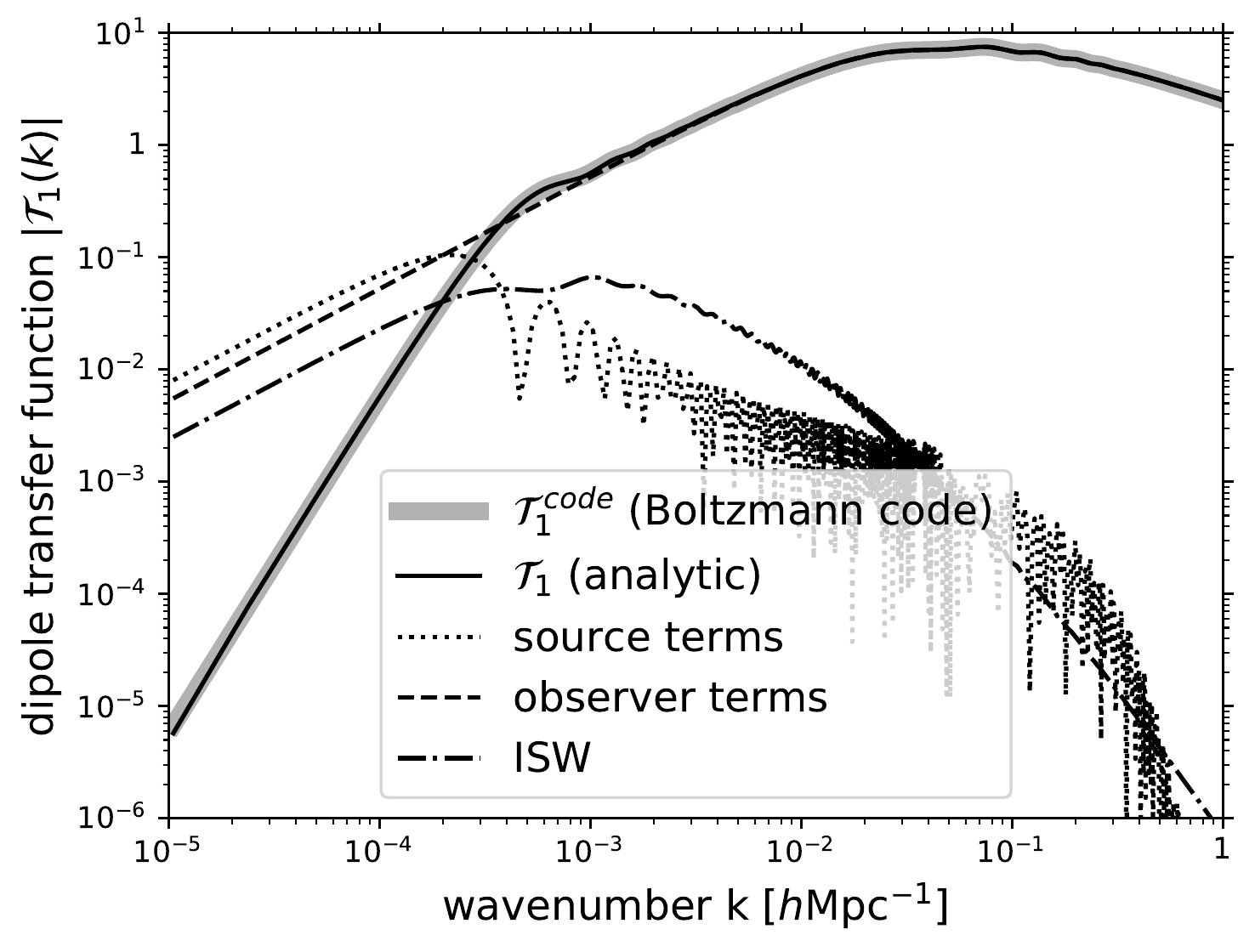}
\caption{Dipole transfer function $\TT_1$ and individual contributions as a function of the wavenumber $k$, in the same format as in Fig. \ref{fig:monopole}.}
\label{fig:dipole}
\end{figure}

\subsection{Dipole}
\label{subsec:dipole}
The transfer function of the dipole in the conformal Newtonian gauge is
\bear
\TT_1(k)&=&\left[\TTT_{\Theta}(\eta_*,k)
+\TTT_{\psi}(\eta_*,k)\right]j_1(k\bar{r}_*)  \nnn
&&
+k \TTT_{v_\gamma }(\eta_*,k)j_1'(k\bar{r}_*)  
 -\frac{k}{3} \TTT_{v_m }(\eta_0,k) \nnn
&&
+\int_0^{\bar{r}_*} d\bar{r}\,\left[\TTT'_{\psi}(\eta,k)
 -\TTT'_{\phi}(\eta,k)\right] j_1(k\bar r)~.
\label{eqn:transfer_function_dipole}
\enar
Similar to the monopole, the dipole transfer function has the same
contributions of the source terms (the photon temperature 
fluctuation~$\Theta_*$, the gravitational redshift~$\psi_*$, and
the baryon-photon velocity~$v_{\gamma*}$) and the integrated Sachs-Wolfe
effect. These contributions are already shown in 
Fig.~\ref{fig:MonopoleContributions}, but they are multiplied 
with the spherical Bessel function for the dipole ($l=1$: gray dashed).
The key difference compared to the monopole transfer function arises
from the observer terms. The observer velocity~$v_m$ contributes to the
dipole via the Doppler effect, whereas the gravitational redshift~$\psi_o$ and the coordinate
lapse~$\delta\eta_o$ drop out in the dipole transfer function.

Figure~\ref{fig:dipole} plots the dipole transfer function~$\TT_1(k)$
and its individual contributions. The observer velocity (dashed) is
the dominant contribution on all scales, and this contribution ($-kv_m/3$) is positive
as shown in Fig.~\ref{fig:MonopoleContributions} (but note that the dotted
curve is $kv_m/30$ instead of $kv_m/3$ with another factor ten to fit in the plot).
On large scales $k<10^{-3}\hmpc$, the source terms (dotted)
and the integrated Sachs-Wolfe term (dot-dashed) become comparable to
the observer velocity contribution (dashed), and these contributions
cancel each other to yield the dipole transfer function (solid) in
proportion to~$k^3$. As evident in Fig.~\ref{fig:MonopoleContributions}, 
the individual transfer functions such as $\TTT_{\Theta}$,
$\TTT_\psi$, and so on become constant on large scales, and hence 
they all fall as~$k$ due to the spherical Bessel function $j_1(x)$ (or due to the
extra~$k$-factor for~$\TTT_{v_\gamma}$ and~$\TTT_{v_m}$). Upon cancellation
of these contributions on large scales, the dipole transfer function picks
up extra~$k^2$ (as in the monopole transfer function) to scale with~$k^3$.

Again to check the validity of our analytical expression of the dipole
transfer function, we plot the dipole transfer function~$\TT_1^{\rm code}(k)$
(gray solid) from the Boltzmann codes. 
The numerical computation in the Boltzmann codes is performed
in the (dark-matter) comoving-synchronous gauge, where the matter velocity
is zero ($v_m=0$). Since the dipole is the spatial energy flux of the CMB photon
distribution measured by the observer, the dipole transfer function is literally the relative
velocity between the observer and the CMB photon fluid (see Appendix \ref{app:SynchronousGauge}; this point was also emphasized in \cite{Zibin_Scott_2008}).
Therefore, in the comoving-synchronous gauge, the dipole transfer function can be obtained as
\beeq
\TT_1^{\rm code}(k)=\frac{k}{3}\TTT^{\rm sync}_{v_\gamma}(\eta_0,k)~.
\eneq
As in the case of the monopole transfer function, our analytical expression provides an accurate description of the dipole transfer function, in particular on large scales, where the cancellation of the individual contributions takes place. The dipole transfer function is again dominated by the contribution at the observer position on small scales, where our approximation is expected to be less accurate.

Using the dipole transfer function, we numerically compute the dipole power
in eq.~\eqref{eqn:definition_Cl}
\beeq
C_1=\frac13\sum_m\left\langle|\hat a_{1m}|^2\right\rangle=4.51\times10^{-6}
~,
\eneq
and the rms fluctuation of the relative velocity
\begin{equation}
\begin{split}
\sigma^2_{v_r}&=\left\langle \vec v_r\cdot \vec v_r\right\rangle=
\int d\ln k~\Delta_\zeta^2(k)|k\TTT_{v_r}(\eta_0,k)|^2 \\\
&= \frac{9 C_1}{4\pi}=\left(540 \, \text{km/s} \right)^2
\, .
\end{split}
\end{equation}
The Planck measurement \cite{2020A&A...641A...1P} of the CMB dipole anisotropy yields that our rest frame is moving $369.82\pm0.11$~km/s with respect 
to the CMB rest frame, consistent with our theoretical expectation of the one-dimensional rms relative velocity $\sigma_{v_r}/\sqrt{3}=311$~km/s. 
Note that our scalar velocity potential $v_\gamma$ is related
to the variable $\theta_\gamma$ in the convention of \cite{1995_Ma_Bertschinger} through $\theta_\gamma=k^2v_\gamma$.

It should be emphasized that the dipole is a measure of the relative velocity, not the absolute
velocity. The velocity potential $v$ gauge-transforms as $\tilde{v}=v-\xi$, in the same way for all species,
such that the relative velocity is gauge-invariant. In relativity, the absolute velocity has no physical meaning
and only the relative velocity at the same spacetime position has physical significance. At different positions,
not only the observer velocity but also the photon velocity vary from those at our position, invalidating the
notion that the CMB rest frame provides an absolute frame for all observers in the Universe.

\subsection{Large-scale limit of the monopole and the dipole transfer functions}
\label{subsec:large_scale_limit}
Having presented our numerical calculations of the monopole and the dipole
transfer functions in eqs.~\eqref{eqn:transfer_function_monopole}
and~\eqref{eqn:transfer_function_dipole},
here we investigate their large-scale behavior 
analytically by taking the limit $k\RA0$. For a small argument
$x=kr_*\ll1$, the spherical Bessel function can be approximated as
\begin{equation}
j_l(x)= \frac{2^l \,l!}{(2l+1)!}\, x^l + \mathcal{O}(x^{l+2})\, ,
\label{eqn:SphericalBesselFunctionSmallArgument}
\end{equation}
and for the monopole and the dipole, they become
\beeq
j_0(x)\simeq 1+{\cal O}(x^2)~,\qquad\qquad
j_1(x)\simeq \frac13x+{\cal O}(x^3)~.
\eneq

In the limit $k\RA0$,
the  monopole transfer function in eq.~\eqref{eqn:transfer_function_monopole}
is approximated as
\bear
\begin{split}
\TT_0(k)\simeq&\, \TTT_{\Theta}(\eta_*,k)
+\TTT_{\psi}(\eta_*,k) -\TTT_{\psi}(\eta_o,k) \\\
&+H_o\TTT_{v_m}(\eta_o,k) 
+\int_0^{\bar{r}_*} d\bar{r}\,\left[\TTT'_{\psi}(\eta,k)
 -\TTT'_{\phi}(\eta,k)\right]~,
 \label{eqn:monopole_LSL}
\end{split}
\enar
where the baryon-photon velocity~$v_{\gamma*}$ term is dropped due to its
$k$-factor and the spherical Bessel function~$j_0'(x)$. 
Figure~\ref{fig:MonopoleContributions} shows all these contributions are constant in $k$ 
on large scales. In this large-scale
limit, the integrated Sachs-Wolfe term can be analytically integrated by part
to yield
\beeq
\TT_0(k)\simeq \TTT_{\Theta}(\eta_*,k) 
+\TTT_{\phi}(\eta_*,k) 
+H_o\TTT_{v_m}(\eta_o,k) - \TTT_{\phi}(\eta_o,k) ~,
\eneq
where $d\bar r$ is the line-of-sight integration ($d/d\bar r =\pa_r-\pa_\eta$) and
the two gravitational redshift terms~$\psi$ are canceled.
The conservation equation
of the photon energy-density takes the form 
\beeq
\dot\Theta+\dot\phi=0~,
\label{eqn:photon_energy_density_cons}
\eneq
on large scales (see, e.g., \cite{1995_Ma_Bertschinger}), yielding
\beeq
\Theta(\eta_*)+\phi(\eta_*)=
\Theta(\eta_0)+\phi(\eta_0)=
\mathfrak{C}~,
\eneq
where $\mathfrak{C}$ is a constant
and we suppressed the scale dependence as these quantities are taken
in the limit $k\RA0$. This further simplifies the monopole transfer 
function as
\beeq
\TT_0(k)\simeq \TTT_{\Theta}(\eta_0,k)+H_o\TTT_{v_m}(\eta_o,k)~.
\eneq

Combining the time-time and the time-space components of the Einstein equation,
we derive the relation between the velocity potential and the matter density
on large scales:
\beeq
v_m=-{\delta\rho_m\over3aH\bar\rho_m}={\delta\rho_m\over a\dot{\bar\rho}_m}~,
\label{eqn:v_m}
\eneq
where we used the background conservation equation for the matter density.
Further assuming the adiabaticity for each species at $k=0$
\beeq
{\delta\rho_m\over\dot{\bar\rho}_m}={\delta\rho_\gamma\over
\dot{\bar\rho}_\gamma}~,
\label{eqn:adiabaticity}
\eneq
the two contributions to the monopole transfer function~$\TT_0(k)$ cancel
in the large-scale limit.
The next-leading order contribution ($\propto k$) could come from
the same terms in eq. \eqref{eqn:monopole_LSL}, as the next-leading order in~$j_0(x)$
is proportional to~$k^2$. Each term in eq. \eqref{eqn:monopole_LSL} can be expanded
as a power series in $k$, and the subsequent derivations are exactly the same for terms
in proportion to $k$, since the conservation eq. \eqref{eqn:photon_energy_density_cons}
is valid up to $k^2$.
While we chose the observer moving together with matter for computing $\delta\eta_o$,
the coordinate lapse $\delta\eta_o$ is in fact independent of this choice, as discussed in
Section~\ref{subsec:GI_background_temp}.

The absence of contributions that are independent of scales or in proportion
to~$k$ is the consequence of the equivalence principle, which states the
equality of the gravitational and the inertial mass. Consequently,
a local observer cannot tell the existence of a uniform gravitational 
force, as the reference frame and the apparatus of the local observer
are affected altogether in the same way.
For our calculations, a uniform gravitational potential corresponds
to the constant contributions in individual
transfer functions such as~$\TTT_\psi$, while
a uniform gravitational acceleration corresponds to the contributions
in proportion to~$k$ (or the gradient of the potential contributions).
It was shown \cite{2012PhRvD..85b3504J,Biern_2017,Scaccabarozzi_2018,grimm2020galaxy} that
the theoretical descriptions of the luminosity distance and galaxy clustering
are devoid of such contributions.

Applying the equivalence principle to the CMB monopole transfer 
function~$\TT_0(k)$, we find that the gravitational potential fluctuations 
or the gravitational accelerations of wavelength larger than $\bar{r}_*$
act as uniform fields and they have no impact on our local physical observables
such as the observed CMB temperature (or the monopole fluctuation).
This argument is borne out by the cancellation of individually diverging
contributions to the monopole power at low~$k$ in our numerical and analytical calculations of
the monopole transfer function~$\TT_0(k)$. Keep in mind that if any of the
potential contributions such as the coordinate lapse~$\delta\eta_o$ 
at the observer position is ignored, the cancellation of each contribution
at low~$k$ would not take place, and the monopole transfer function~$\TT_0(k)$
 is non-vanishing even in the limit $k\RA0$, leading to an infinite
monopole power~$C_0$ (hence the observed CMB temperature) or the monopole 
power highly sensitive to the lower cut-off scale in the integral
of eq.~\eqref{eqn:definition_Cl}.  This would work against the equivalence
principle, as the super-horizon scale fluctuations dictate our local 
observables or put it differently we can infer the existence of
such super-horizon scale
fluctuations based on our local observables.

In the limit $k\RA0$, the dipole transfer function in 
eq.~\eqref{eqn:transfer_function_dipole} is approximated as
\bear
\TT_1(k)&\simeq&\left[\TTT_{\Theta}(\eta_*,k)
+\TTT_{\psi}(\eta_*,k)\right]{k\bar{r}_*\over3} +{k\over3}
\TTT_{v_\gamma }(\eta_*,k) \\
&&
 -\frac{k}{3} \TTT_{v_m }(\eta_0,k) 
+\int_0^{\bar{r}_*} d\bar{r}\,\left[\TTT'_{\psi}(\eta,k)
 -\TTT'_{\phi}(\eta,k)\right] {k\bar r\over3}~.~~~\nonumber
\enar
Given that the individual transfer functions are constant at low~$k$,
it is apparent that the dipole transfer function goes at least as~$k$.
Using the conservation equation~\eqref{eqn:photon_energy_density_cons},
we first replace the term~$\phi'$ with~$\Theta'$ in the integrated
Sachs-Wolfe term, and by performing the integration by part, the
dipole transfer function can be expressed as 
\bear
\frac3k\TT_1(k)&\simeq& \TTT_{v_\gamma }(\eta_*,k)
 - \TTT_{v_m }(\eta_0,k) \label{eqn:LSL_dipole} \\
 &&
+ \int_0^{\bar{r}_*} d\bar{r}\,\left[\TTT_{\psi}(\eta,k)
 +\TTT_{\Theta}(\eta,k)\right]~.~~~\nonumber
\enar
Similarly to the monopole, we make use of the conservation equation for
the photon  energy-momentum on large scales
\beeq
\dot v_\gamma = \frac1a\psi+{1\over a(\bar\rho_\gamma+\bar p_\gamma)}\delta
p_\gamma=\frac1a\left(\psi+\Theta\right)~, 
\eneq
and integrate along the line-of-sight to obtain
\beeq
\TTT_{v_\gamma}(\eta_*)-\TTT_{v_\gamma}(\eta_0)
= -\int_0^{\bar{r}_*} d\bar r \left(\TTT_{\psi}+\TTT_{\Theta}\right)~,
\eneq
where we suppressed the scale dependence.  Using the adiabaticity condition
in eq.~\eqref{eqn:adiabaticity} and the velocity potential for matter in eq.~\eqref{eqn:v_m},
we find that the dipole transfer function vanishes on large scales.
Again, the large-scale limit of the conservation equation is valid up to~$k^2$,
and the leading term in the dipole transfer function comes in proportion
to~$k^3$.

\begin{figure}
\includegraphics[width=\linewidth]{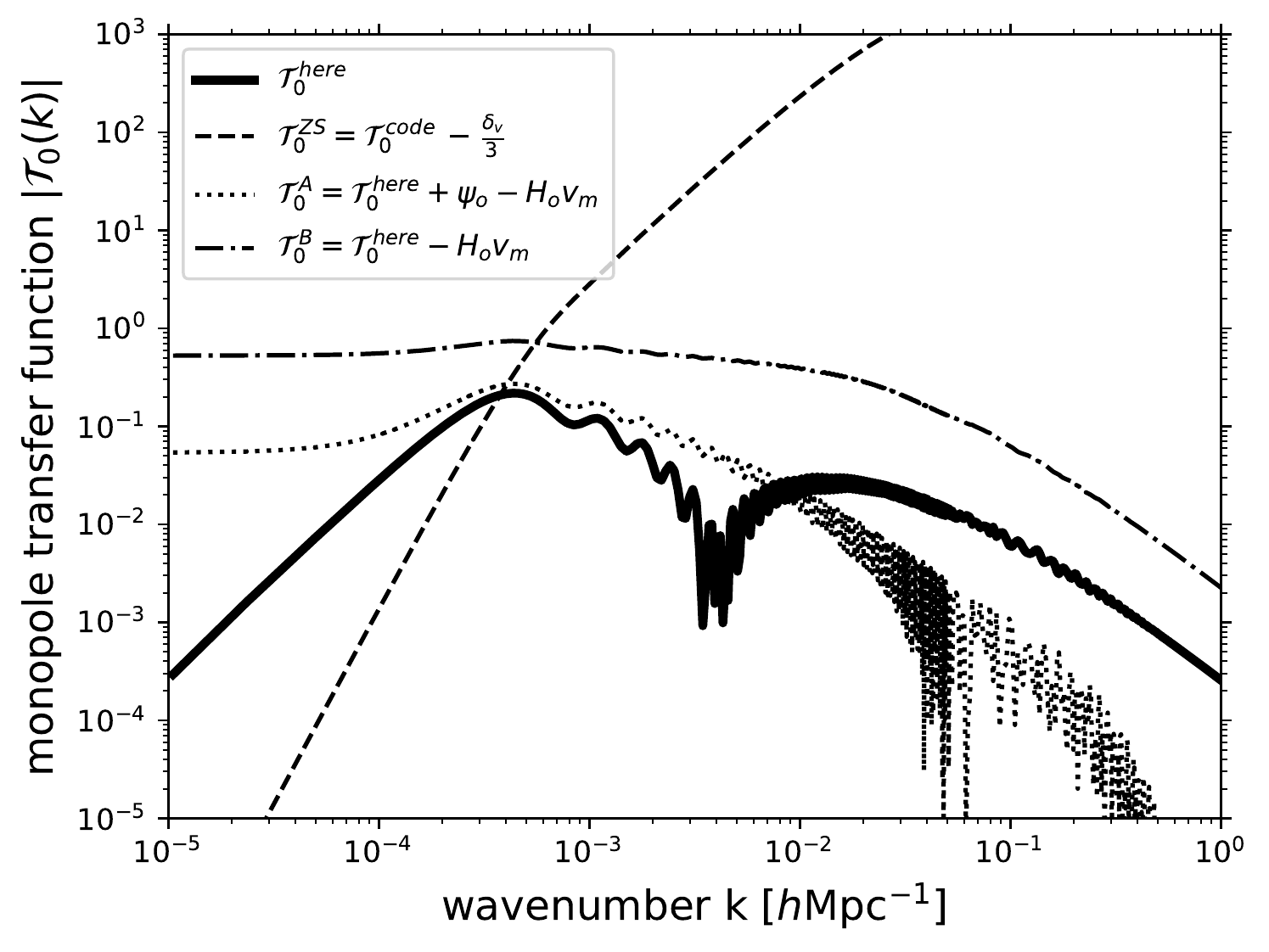}
\caption{Comparison of the analytic expression of the monopole transfer function derived in this paper (solid) and the one derived in \cite{Zibin_Scott_2008} (dashed), the one without the observer terms (dotted), and the one without the coordinate lapse (dot-dashed). All the monopole transfer functions except 
$\TT_0^\text{\tiny{here}}$ lead to an infinite monopole power.}
\label{fig:MonopoleComparison}
\end{figure}

\subsection{Comparison to previous work}
\label{subsec:comparison}

\subsubsection{Zibin \& Scott 2008}
In the comprehensive work by Zibin \& Scott \cite{Zibin_Scott_2008}, 
an analytic expression
of the observed CMB temperature anisotropies was derived with close 
attention to the gauge issues associated with the monopole and the dipole
transfer functions. Despite the apparent difference in their approach 
and the notation convention, the theoretical description in their
work is based on the same assumptions adopted in this work --- CMB photons decouple abruptly 
from the local plasma at the temperature~$T_E$ (E:~emission)
in the rest frame of the baryon-photon plasma
with four-velocity~$u^\mu_E$, and they freely propagate to the observer,
moving together with matter,
where the photons are received with temperature~$T_R$ (R:~reception)
along the observed direction~$n^\alpha$.

Compared to our expression in 
eq.~\eqref{eqn:analytical_expression}, the key difference lies in the
observer position today. Here we briefly compare how the difference arises
in the work \cite{Zibin_Scott_2008}. The main calculation is to derive
the exact description of the temperature ratio  in their eq.~(24)
and its linearized equation~(29):
\beeq
\int_R^Edt~H_nN=\int_{\bar t_R}^{\bar t_E}dt
~H_nN+\left(\bar H\delta t_D\right)^E_R~.
\eneq
where $H_n$ is the derivative of the photon frequency along the line-of-sight,
$N$ is the time lapse in the ADM formalism \cite{1962rdgr.book..127A}, and $\delta t_D$ describes
the deviation of the exact positions at emission and reception from the
background coordinate~$\bar t$.
The left-hand side is essentially the line-of-sight integration, but replaced
with the coordinate integration to yield the frequency ratio of photons
emitted at the position~$E$ and received at the position~$R$. Since the
computation of the line-of-sight integration is performed for an ADM normal
observer with~$u^\mu_\text{\tiny{ADM}}$, 
they perform Lorentz boosts~$\delta t_B$
both at the emission and the reception
to match the physical frames of the baryon-photon plasma and the observer,
of which the four-velocity is then 
$u^\mu=u^\mu_\text{\tiny{ADM}}+\delta t_B{}^{,\mu}$.
This yields their main equation~(35) for the observed CMB temperature
anisotropies:
\beeq
{\delta T(n^\mu)\over\bar T_R}=\int_{\bar t_R}^{\bar t_E}dt
~\delta(H_nN)+\left(\bar H\delta t_D+n^\mu\delta t_{B,\mu}\right)^E_R~,
\label{theirs}
\eneq
where the background temperature at reception is defined as
\beeq
\bar T_R:=T_E\exp\left[\int_{\bar t_R}^{\bar t_E}dt~\bar H\right]~,
\label{bgZS}
\eneq
and their photon propagation direction~$n^\mu$ in a coordinate is
the opposite of our observed direction~$n^\alpha$ in addition to the
overall scale factor $a$:
\beeq
n^\mu=-\frac1a\left(0,n^\alpha\right)+{\cal O}(1)~.
\eneq

This formula describes the observed anisotropies~$\hat\Theta(\hat n)$, 
corresponding to our eq.~\eqref{eqn:ThetaObs}, and the derivation
above is equivalent to our eq.~\eqref{eqn:redshift_relation}
for the observed redshift with $T_*$ replaced by~$T_E$. In more detail,
the line-of-sight integration was computed in the conformal Newtonian
gauge in their eq.~(41) as
\beeq
\int_{\bar t_R}^{\bar t_E}dt~\delta(H_nN)\RA
\int_{0}^{\bar r_*}d\bar r~(\psi-\phi)' 
+ \psi\bigg|_o^* ~,
\eneq
where we used our notation convention in the right-hand side.
The boost parameters~$\delta t_B$ are set to transform an ADM normal observer
to an observer without any spatial energy flux in their eq.~(43):
\beeq
\delta t_B =-\frac{H\psi-\dot{\phi}}{4\pi G (\rho+p)} ~.
\eneq
Using the Einstein equation, we derive that the boost parameter is indeed
\beeq
\delta t_B\RA-av_N~,
\eneq
corresponding to an observer velocity, moving together with the total matter 
component, where the subscript~$N$ denotes that the quantity is computed
in the conformal Newtonian gauge, in addition to the two gravitational 
potentials~$\psi$ and~$\phi$.
 Since the radiation energy density is already smaller at 
the decoupling, this velocity at emission
would correspond to the matter component.
Note that the velocity at emission in our formula is the one for the baryon-photon
fluid, regardless of the validity of the approximation in the analytical expression, as the rest frame
of the photon emission is specified by the baryon-photon plasma.
However, we showed in Fig.~\ref{fig:MonopoleContributions} that since the matter
velocity is the same as the baryon-photon velocity on large scales, this difference
has no impact on the monopole transfer function.
At reception, this boost parameter gives the matter velocity, as in our
formula.

Now we come to the temporal displacement terms~$\delta t_D$. Compared to our
eq.~\eqref{eqn:ThetaObs} for~$\hat\Theta(\hat n)$, the absence of
the temperature fluctuation~$\Theta_*$ at decoupling is apparent
in eq.~\eqref{theirs}, as their equations are derived specifically 
for the hypersurface in which the photon density is uniform, 
i.e., $\Theta_*\equiv0$. This corresponds to a gauge choice, set by~$\delta
t_D$ at emission. The time coordinate of the emission point is~$\bar t_E$,
corresponding to our~$\eta_*$, and the CMB photon temperature has no
fluctuation, i.e.,
\beeq
T_E=:\bar T(t_E)=\bar T(\bar t_E)\left(1-H\delta t_D\right)~.
\eneq
By identifying to eq.~\eqref{eqn:decouplingTemperature},
we find that the temporal displacement~$\delta t_D$ at emission
plays a role of restoring the broken gauge symmetry:
\beeq
-H\delta t_D\RA \Theta_*~.
\label{Tsrc}
\eneq

Indeed, the temporal displacement terms~$\delta t_D$ 
were computed at both positions by gauge-transforming to the uniform energy 
density gauge in their eq.~(42)
\beeq
\delta t_D = -{\delta\rho_N\over \dot{\bar{\rho}}} 
= -\frac{3H(H\psi-\dot{\phi}) +
 \frac{\Delta }{a^2}\phi}{12\pi G H(\rho+p)} \, ,
\eneq
where the matter or the photon density fluctuations are also vanishing 
with the adiabaticity assumption. Using the Einstein equation, we find that
the temporal displacements are 
\beeq
\delta t_D\RA{\delta_N\over 3H}~,
\eneq
where $\delta_N:=\delta-\chi(\dot{\bar\rho}/\bar\rho)$ is the gauge-invariant
expression for the density fluctuation in the conformal Newtonian gauge.
While this transformation~$\delta t_D$ correctly captures the emission point
in eq.~\eqref{Tsrc}, the observer position today is not at the hypersurface
of the uniform energy density or the matter density. In fact, it was argued
\cite{Zibin_Scott_2008} that $\delta t_D$ at reception is 
\textit{not uniquely} fixed
by any physical prescription, the choice of~$\delta t_D$ at reception only
affects the monopole, and they chose it to be the gauge-transformation to the uniform energy density gauge as above.

We argued in Sec.~\ref{subsec:GI_background_temp} that the observer position
is uniquely fixed, once we assume that the observer is moving together with
the matter, as its trajectory in time determines the current position
in eqs.~\eqref{lapse} and~\eqref{eqn:DeltaEta}. 
In fact, the coordinate lapse $\delta\eta_o$ in eq.~\eqref{eqn:DeltaEta} is generic for
all observers at the linear order.
By further assuming that
the observer motion is geodesic, the expression for the observer position,
or the coordinate (time) lapse is simplified as $\delta\eta_o=-v_o$
in eq.~\eqref{geodesic}. The temporal displacement~$\delta t_D$ at 
reception in \cite{Zibin_Scott_2008} is then expressed as
\beeq
\delta t_D\RA \delta\eta_o -
\frac{\Delta\phi}{12\pi G a^2H(\rho+p)}\bigg|_o
=\delta\eta_o+{\delta_v\over3H_0}~,
\eneq
where $\delta_v:=\delta_N-av_N(\dot{\bar\rho}/\bar\rho)$ is the gauge-invariant
expression for the matter density in the comoving gauge (or the usual
matter density from the Boltzmann codes) and we used the
Einstein equation in the last equality. 

One subtlety in \cite{Zibin_Scott_2008} is that the ``background'' observed
temperature~$\bar T_R$ in eq.~\eqref{bgZS} or their eq.~(34) needs
further clarification.
In order to set $\bar T_R$ equal to our coordinate-independent
background CMB temperature~$\bar T:=\bar T(\bar\eta_o)$, the explicit
definition of~$\bar t_R$ and~$\bar t_E$ in a coordinate-independent way
would be needed. Upon setting $\bar T_R\RA\bar T$, we derive the
relation of the observed CMB temperature anisotropies 
in \cite{Zibin_Scott_2008} in eq.~\eqref{theirs} to our expression as
\beeq
\hat\Theta_\text{\tiny{ZS}}= \hat\Theta_\text{{here}}
-{1\over3}\delta_v~.
\eneq
We emphasize again that the difference arises due to the observer position
today and it only affects the monopole transfer function.
Figure~\ref{fig:MonopoleComparison} compares the monopole
transfer function in this work (solid) and in \cite{Zibin_Scott_2008} (dashed).
Given that the matter density fluctuation in the comoving gauge $\delta_v \propto k^2\phi$, the monopole transfer function is expected to behave as $\TT_0^\text{\tiny{ZS}}(k)\propto k^2$ on large scales. However, the adiabatic condition in the comoving gauge imposes
\begin{equation}
\Theta=\frac{1}{4}\delta_\gamma=\frac{1}{3}\delta_m \, ,
\end{equation}
leading to another cancellation on large scales, and the resulting behavior is $\TT_0^\text{\tiny{ZS}}(k)\propto k^4$ and the monopole power~$C_0^\text{\tiny{ZS}}$
is devoid of infrared divergences. 
Note that the coordinate lapse vanishes in the comoving gauge and our gauge-invariant expression $\hat{\Theta}$ coincides with $\Theta$ in the comoving gauge.
However, as the
density fluctuation~$\delta_v$ becomes dominant on small scales, 
the monopole transfer function $\TT_0^\text{\tiny{ZS}}(k)\propto \delta_v(k)$
(dashed) increases with~$\simeq k^{0.15}$, leading to the UV divergence
in the monopole power~$C_0$. It was concluded \cite{Zibin_Scott_2008}
that ``this divergence makes it impossible to quantify the total power~$C_0$.''
The dipole and other multipoles are the same as derived in this work. Note
that the transfer function in \cite{Zibin_Scott_2008} is defined with an
extra factor five: $\TT_l^\text{\tiny{ZS}}=5\TT_l^\text{\tiny{here}}$.

\subsubsection{Case A: Without the observer terms}
There exist few work that focus on the gauge-invariance of the full observed CMB 
temperature anisotropies, rather than the expressions for the higher angular multipoles.	
In Hwang \& Noh \cite{1999PhRvD..59f7302H} the gauge-invariance of the observed CMB temperature
anisotropies was investigated, though the main focus is not the monopole anisotropy. 
Starting with the same assumption that
the observed CMB temperature anisotropies at one direction originate from 
the photons emitted at a single point, they derive the expression for the 
observed CMB temperature anisotropies in their eq.~(14):
\bear
&&
\label{HW99}
{T(\hat n)\over \bar T(\eta_o)}-1=\left(\Theta+H\chi\right)_*
+\left({v_\chi}_{,\alpha}n^\alpha+\alpha_\chi\right)_*
-\left({v_\chi}_{,\alpha}n^\alpha\right)_o \nnn
&&
\qquad\qquad
-H_o\chi_o-\alpha_\chi{}_o+\int_0^{\bar{r}_*} \!\!\!
d\bar{r}\left(\alpha_\chi-\varphi_\chi\right)'~,
\enar
where we used our notation convention to express the right-hand side of the
equation. They noted that the first two terms at the source is
the gauge-invariant combination $(\Theta+H\chi)_*\RA\Theta_{\chi*}$.
More importantly, they identified the gauge-dependence of the expression
for the observed CMB temperature anisotropies due to the term~$H_o\chi_o$.
Since it is independent of the angular direction (also the 
term~$\alpha_{\chi o}$), they argued that
it will be absorbed into the background CMB temperature, or the angle average
$\langle T\rangle_\Omega$ in eq.~\eqref{skyavg}. So, they arrived at 
the gauge-invariant expression for the observed CMB temperature anisotropies
in their eq.~(15):
\beeq
\hat\Theta_\text{\tiny{A}}={\Theta_{\chi*}}
+\left({v_\chi}_{,\alpha}n^\alpha+\alpha_\chi\right)_*
-\left({v_\chi}_{,\alpha}n^\alpha\right)_o 
+\int_0^{\bar{r}_*} \!\!\!
d\bar{r}\left(\alpha_\chi-\varphi_\chi\right)'~,
\label{eqn:Theta_A}
\eneq
after removing the two terms at the observer position without the angular 
dependence. Despite the procedure to subtract the contribution to 
$\langle T\rangle_\Omega$, eq.~\eqref{eqn:Theta_A} still has the monopole fluctuation.

In comparison to our gauge-invariant expression in 
eq.~\eqref{eqn:analytical_expression}, it is clear that
the gauge-dependence in eq.~\eqref{HW99} arises from the background temperature
at the observer position~$\bar T(\eta_o)$ in the left-hand side. As shown in 
eq.~\eqref{eqn:RelationBackgroundTemperature}, the value of $\bar T(\eta_o)$ 
depends on our choice of coordinate, and it is composed of the background 
temperature~$\bar T$ in a homogeneous universe and the coordinate 
lapse~$\delta\eta_o$. This yields extra terms in eq.~\eqref{HW99} that
are not included in~$\hat\Theta_\text{\tiny{A}}$:
\beeq
-H_o\delta\eta_o-H_o\chi_o-\alpha_{\chi o}~.
\eneq
Consequently, we derive the relation of the observed CMB temperature 
anisotropies to our expression as
\beeq
\hat\Theta_\text{\tiny{A}}=\hat\Theta_\text{\tiny{here}}+\alpha_{\chi o}
-H_ov_{\chi o}~.
\eneq
Figure~\ref{fig:MonopoleComparison} shows the monopole 
transfer function for~$\hat\Theta_\text{\tiny{A}}$ (dotted).
With two extra potential terms missing, no cancellation occurs on large
scales, and the transfer function does not vanish in the infrared. Therefore, despite the absence of
divergences in the UV, the monopole power is also infinite $C_0^\text{\tiny{A}}=\infty$.

\subsubsection{Case B: Without the coordinate lapse}
Another common case in literature is to neglect the coordinate lapse. Though 
calculations are done properly (up to $\delta\eta_o$), less attention is paid to
the gauge-invariance of the resulting expression.
In the pioneering work \cite{Sachs_Wolfe_1967}, the expression for the
observed CMB temperature anisotropies was derived (see also the following
work \cite{1984PhLB..135..279A,1994PhRvD..49.1845S}) by using the
relation of the observed CMB temperature
to the observed redshift in eq.~\eqref{eqn:redshift_relation}.
Perturbing all the quantities in eq.~\eqref{eqn:redshift_relation}, 
we obtain their key equation for the observed CMB temperature anisotropies
\beeq
{\Delta T(\hat n)\over T(\hat n)}
={\Delta T_*\over T_*}-{\Delta z\over1+z_{\rm obs}}~,
\eneq
similar to our eq.~\eqref{eqn:ThetaObs}. By defining
the last-scattering surface as a hypersurface of the uniform energy density
$\Delta T_*\equiv0$, they computed the perturbation to the observed redshift.
However, the devils are again in details, and the ambiguities arise from
identifying the correct background quantities. At the linear order in 
perturbations, $\Delta z$ can be safely equated to
$(1+z_*)\delta z$ in our notation convention, but without the coordinate
lapse~$\delta\eta_o$. Again, the missing lapse term owes to the lack of
proper consideration of the observer position today. The final expression
for the observed CMB temperature anisotropies is then
\beeq
\hat\Theta_\text{\tiny{B}}=\hat\Theta_\text{\tiny{here}}-H_ov_{\chi o}~.
\eneq
Figure~\ref{fig:MonopoleComparison} shows the monopole 
transfer function for~$\hat\Theta_\text{\tiny{B}}$ (dot-dashed).
Similar to~$\hat\Theta_\text{\tiny{A}}$, the monopole power is plagued
with infrared divergences, but without UV divergences.
The monopole power~$C_0^\text{\tiny{B}}$ is again infinite.

\section{Discussion}
\label{sec:discussion}
We have derived an analytic expression for the observed CMB temperature
anisotropies~$\hat\Theta(\hat n)$
in eqs.~\eqref{eqn:ThetaObs} and~\eqref{eqn:analytical_expression}
and investigated the gauge issues in comparison to previous work.
It is well-known that a general coordinate transformation in 
eq.~\eqref{eqn:GeneralCoordinateTransformation}
induces a gauge transformation for the temperature fluctuation~$\Theta$
at the observer position~$x^\mu_o$:
\beeq
\tilde \Theta(\tilde x_o^\mu)=\Theta(x_o^\mu)+{\cal H}_o\xi_o~,
\label{gdmono}
\eneq
as the background CMB temperature depends only on the time coordinate
\beeq
\bar T(\tilde\eta_o)=\bar T(\eta_o)+\bar T'(\eta_o)\xi_o~.
\label{gdT}
\eneq
Consequently, the temperature fluctuation~$\Theta$ at the observer position
is gauge-dependent (in fact at any position).
However, since the gauge mode is independent of observed
angular direction~$\hat n$, only the monopole fluctuation~$\Theta_0$ 
is gauge-dependent, 
and the other angular multipoles~$\Theta_l$ with $l\geq1$ 
are gauge-invariant, when decomposed
in terms of observed angle~$\hat n$.
This statement is correct, but the gauge-dependence of
the temperature fluctuation~$\Theta$ at the observer position indicates
that it cannot be the correct description of the observed CMB temperature
anisotropies. To put it differently, the theoretical prediction for the observed values of the CMB
temperature anisotropies should be independent of our choice of gauge 
condition.

With a few exceptions \cite{1999PhRvD..59f7302H,Zibin_Scott_2008}, relatively
little attention has been paid in literature to this flaw in the theoretical
description of the observed CMB temperature anisotropies, largely because
the angular multipoles~$a_{lm}$ of the temperature fluctuation~$\Theta(\hat n)$
with $l\geq1$ are gauge-invariant and they contain most
of the cosmological information, and also because
the angle-averaged CMB temperature, or the combination of the background
and the fluctuation
\beeq
\left\langle T\right\rangle_\Omega:=\int{d^2\hat n\over4\pi}~T(\hat n)
=\bar T(\eta_o)\left[1+\Theta_0(x^\mu_o)\right]~,
\label{eqn:angleaverageGD}
\eneq
is well measured and gauge-invariant,
where $\Theta_0(x^\mu_o)$ is the angle average (or monopole) of the temperature
fluctuation~$\Theta(x^\mu_o)$ at the observer position and it gauge
transforms as in eq.~\eqref{gdmono}. However, we showed that the monopole
power~$C_0$ computed by using the gauge-dependent description 
of~$\Theta(\hat n)$ is {\it infinite}, logarithmically diverging in 
the infrared.

The angle-averaged CMB temperature is expected to fluctuate around 
the background temperature from place to place, but its variance~$C_0$ cannot 
be infinite. Figures~\ref{fig:monopole} and~\ref{fig:MonopoleContributions}
show that the infrared divergence in the monopole power originates from
the gravitational potential contributions on very large scales or low~$k$.
Those contributions act as a uniform gravitational potential on scales
smaller than their wavelength, and they should have {\it no} impact on
any local measurements, as any test particles and the measurement
apparatus would move together, according to the equivalence principle.
In fact, our investigation of the monopole fluctuation in the large-scale limit
in Sec.~\ref{subsec:large_scale_limit} proves that our gauge-invariant
expression for the observed CMB temperature anisotropies~$\hat\Theta$ 
in eqs.~\eqref{eqn:ThetaObs} and~\eqref{eqn:analytical_expression}
(as opposed to the gauge-dependent expression~$\Theta$) indeed contains
numerous components that act as a uniform gravitational potential on very
large scales, but their large-scale contributions cancel to yield the
leading-order contribution in proportion to~$k^2$. The infrared divergence
of the monopole power arises due to the use of the gauge-dependent
expression of the CMB temperature fluctuation, which neglects one or a few
contributions, breaking the gauge-invariance and the subtle balance
for the cancellation.  In fact, the cancellation of such contributions
is stronger, as the equivalence principle states that a uniform gravitational 
acceleration cannot be measured locally, preventing any gradient
contributions of the gravitational potentials on large scales.

In Sec.~\ref{subsec:comparison}, we have compared our gauge-invariant
expression for the observed CMB temperature anisotropies to previous
work. The major physical process of the CMB temperature anisotropy formation
was fully identified in the pioneering work by Sachs \& Wolfe in 1967
\cite{Sachs_Wolfe_1967} --- once the baryon-photon plasma cools to decouple,
the CMB photons propagate freely in space, and they are measured by the
observer. This physical process naturally involves the physical quantities
along the photon path as well as those at the decoupling position and 
the observer position. The first contribution is referred to as the integrated
Sachs-Wolfe effect, and the contributions at the source position are
made of the gravitational redshift (or Sachs-Wolfe effect), the Doppler
effect, and the (intrinsic) temperature fluctuation,
while the contributions at the observer position were often neglected
in literature. 
In the comprehensive work by Zibin \& Scott \cite{Zibin_Scott_2008},
this physical process was carefully examined in close attention to the
gauge-invariance of the theoretical description of the observed CMB temperature
anisotropies. The observer should be moving together with baryon and matter
components, at least in the linear-order description, and this observer
motion contributes to the dipole anisotropy.
According to \cite{Zibin_Scott_2008}, however, there still remains one
ambiguity, which is the choice of the observer hypersurface, or the 
time coordinate of the observer position. 
This is evident in eqs.~\eqref{gdmono} and~\eqref{gdT}, and such ambiguity
in defining the observer position is one source for the gauge-dependence
of the analytical expressions in literature. A hypersurface of the uniform
energy density was chosen in \cite{Zibin_Scott_2008}, though it was also noted that this choice is not unique. By specifying the 
hypersurface, their expression for the observed CMB temperature
is gauge-invariant, but the resulting monopole power is UV divergent.

In this work we have shown that there is {\it no} ambiguity in describing
the observed CMB temperature anisotropies, as expected for any physical
observations; the observer position today is
uniquely determined, once a physical choice of the observer is made. The
observer is moving together with the matter component, and its time 
coordinate~$\eta_o$ today can be computed by following the motion of the
observer in time. Compared to the reference time coordinate~$\bar\eta_o$
of the observer in the background in eq.~\eqref{referencetime},
the observer time coordinate today deviates from~$\bar\eta_o$
by the coordinate lapse~$\delta\eta_o$, defined as in eq.~\eqref{lapse}:
$\eta_o:=\bar\eta_o+\delta\eta_o$.
The coordinate lapse~$\delta\eta_o$, of course, gauge-transforms, as the
time coordinate of the observer depends on the choice of coordinate system.
However, it is this extra gauge-dependent term that compensates for the
gauge-dependence in the temperature fluctuation~$\Theta$ at the observer
position, as expressed in eq.~\eqref{eqn:ThetaObs}.
Due to the absence of the angular dependence, the expressions for the
higher-order multipoles are gauge-invariant.

Using our gauge-invariant expression for the observed CMB temperature
anisotropies in eq.~\eqref{eqn:ThetaObs} and the Boltzmann codes
\textsc{class} and~\textsc{camb}, we have numerically computed the monopole
power \textit{for the first time}
\beeq
C_0=1.66\times 10^{-9}~,
\eneq
corresponding to the rms monopole fluctuation
\beeq
\sqrt{\left\langle\hat\Theta^2_0\right\rangle}=\sqrt{C_0\over4\pi}=
1.15\times 10^{-5}~.
\eneq
The largest contribution to the monopole power is the gravitational potential
at the decoupling point, which is also the main source for the Sachs-Wolfe
plateau at low angular multipoles. A finite value of the monopole power 
indicates that the angle-average of the CMB temperature fluctuates in space
around the background CMB temperature $\bar T(\bar\eta_o)$ evaluated
at the reference time coordinate~$\bar\eta_o$, and the COBE FIRAS
measurement~$\langle T\rangle_\Omega$ is {\it not} the background 
temperature~$\bar T(\bar\eta_o)$, but one with the monopole 
fluctuation~$\hat\Theta_0$ at our position. Given that the current measurement
uncertainty is $\sim10$ times larger than the rms fluctuation, the impact
of properly accounting for the difference between~$\bar T(\bar\eta_o)$
and $\langle T\rangle_\Omega$ is negligible 
\cite{Yoo_Mitsou_Dirian_Durrer_2019} for the CMB power spectrum analysis.
With $\bar{T}:=\bar T(\bar{\eta}_o)$ being one of the fundamental cosmological parameters, the monopole fluctuation~$\hat\Theta_0$ (or the angle-averaged anisotropy) can be inferred from the measurement of $\langle T\rangle_\Omega$ once a cosmological model is chosen. This is in contrast to the coordinate-dependent $\bar T(\eta_o)$ and $\Theta_0(x^\mu_o)$ in eq.~\eqref{eqn:angleaverageGD}, which are not separable from the measurement of $\langle T\rangle_\Omega$ due to their ambiguous definition. Hence the monopole fluctuation~$\hat\Theta_0$ is a model-dependent, but coordinate-independent ``observable''.

\acknowledgments
We thank Nastassia Grimm, Ermis Mitsou, Douglas Scott and James Zibin for useful discussions.
We acknowledge support by 
the Swiss National Science Foundation (SNF PP00P2\_176996).
J. Y. is further supported by a Consolidator Grant of the European Research 
Council (ERC-2015-CoG Grant No. 680886).

\appendix
\section{Metric convention and gauge transformation properties of the perturbation variables}
\label{app:metric}
In this Section we introduce the notation convention used in this paper.
The background universe is described by a Robertson-Walker metric:
\begin{equation*}
ds^2 = -a^2(\eta) d\eta^2 + a^2(\eta)\bar{g}_{\alpha\beta}dx^\alpha dx^\beta \, ,
\end{equation*} 
with conformal time $\eta$ and scale factor $a(\eta)$. To account for the inhomogeneities in the universe, we introduce scalar perturbations to the metric tensor with the following convention:
\begin{equation}
\begin{gathered}
g_{00}:=-a^2(1+2\alpha)\, , \qquad
g_{0\alpha}:=-a^2 \beta_{,\alpha}\, , \\\
g_{\alpha\beta}:= a^2\left[(1+2\varphi)\bar{g}_{\alpha\beta} + 2\gamma_{,\alpha\mid\beta} \right]\, .
\end{gathered}
\label{eqn:appendixMetric}
\end{equation} 
The comma represents the coordinate derivative, while the vertical bar represents the covariant derivative with respect to the 3-metric $\bar{g}_{\alpha\beta}$. We do not consider vector and tensor perturbations in this paper. 
The time-like four-velocity vector is introduced as
\begin{equation}
\begin{split}
u^\mu&=\frac{1}{a}\left(1-\alpha, \, U^\alpha\right)\, , \qquad -1 = u_\mu u^\mu \, ,
\end{split}
\end{equation}
and we define the scalar velocity potentials 
\begin{equation}
U^\alpha := -U^{,\alpha} \, , \qquad v:= U+\beta \, .
\end{equation}

Under the general coordinate transformation in eq.~\eqref{eqn:GeneralCoordinateTransformation}, the metric tensor gauge-transforms as
\begin{equation*}
\delta_\xi g_{\mu\nu} =\tilde{g}_{\mu\nu}(x) - g_{\mu\nu}(x) = -\pounds_\xi g_{\mu\nu} \, ,
\end{equation*}
at linear order. The Lie derivative $\pounds$ of a rank 2 tensor is given by 
\begin{equation*}
\pounds_\xi g_{\mu\nu} = g_{\mu\nu,\rho}\,\xi^\rho + g_{\rho\nu}\,\xi^\rho_{\, \, ,\mu} + g_{\mu\rho}\,\xi^\rho_{\, \, ,\nu} \, .
\end{equation*}
For the time-time component, this leads to
\begin{equation*}
\tilde{\alpha}(x)= \alpha(x) -\frac{1}{a}\left(a\xi\right)' \, ,   
\end{equation*}
where the prime denotes the partial derivative with respect to conformal time $\eta$.
The time-space component and the space-space component of the metric reveal the transformation properties of the other scalar perturbation variables introduced in eq.~\eqref{eqn:appendixMetric}: 
\begin{equation*}
\tilde{\beta}=\beta+L'-\xi \, , \qquad \tilde{\varphi}=\varphi-\mathcal{H}\xi \, , \qquad \tilde{\gamma}=\gamma-L \, .
\end{equation*}
Since theoretical descriptions of observables have to be gauge-invariant, it is most convenient to work with gauge-invariant variables \cite{1980PhRvD..22.1882B}:
\begin{equation*}
\alpha_\chi:=\alpha-\frac{1}{a}\chi' \, , \qquad
\varphi_\chi:=\varphi-H\chi \, , 
\end{equation*}
with the scalar shear of the normal observer
\begin{equation*}
\chi:= a(\beta+\gamma') \, ,
\end{equation*}
introduced in \cite{1988cpp..conf....1B}. It gauge-transforms as $\tilde{\chi}=\chi-a\xi$. The two potentials correspond to the Bardeen variables $\alpha_\chi\rightarrow\Phi_A$ and $\varphi_\chi\rightarrow\Phi_H$ in \cite{1980PhRvD..22.1882B}.
The energy-momentum tensor transforms in the same way as the metric tensor. The time-time and the time-space component reveal the transformation properties of the density fluctuation and the scalar velocity potential: 
\begin{equation*}
\tilde{\delta}=\delta-\frac{\bar{\rho}'}{\bar{\rho}} \xi~, \qquad 
\tilde{v}=v-\xi \, .
\end{equation*} 
Therefore, we construct two gauge-invariant quantities
\begin{equation*}
\Theta_\chi:=\frac{1}{4}\delta_\chi^\gamma=\frac{1}{4}\delta_\gamma+H\chi~, \qquad
v_\chi:=v-\frac{\chi}{a}~.
\end{equation*}

\section{Numerical computation of the monopole and dipole transfer functions}
\label{app:SynchronousGauge}
For the black-body radiation, the temperature anisotropy~$\Theta$ is related to the
fluctuation in the photon distribution function~$f$ at the linear order as
\begin{equation}
f(q,\hat n)=-\frac{d\bar{f}}{d \ln q}\Theta(\hat n) \, , 
\end{equation}
where $q$ is the comoving momentum and the photon distribution
function in the background is
\beeq
\bar{f}(q)=\left[\exp\left(\frac{q}{a\bar{T}(\eta)}\right)-1\right]^{-1} \, .
\eneq
Noting that the temperature anisotropy is independent of~$q$, we can
integrate over~$q$ to derive
\beeq
\Theta(\hat n)={2\pi\over a^4\bar\rho_\gamma}\int dq~q^3f(q,\hat n)~,
\eneq
where the photon energy density is
\beeq
\bar \rho_\gamma(\eta)={2\over a^4}\int d^3q~q\bar f(q)~.
\label{bgg}
\eneq

The temperature anisotropy can be decomposed in terms of observed angle,
and the multipole coefficients in eq.~\eqref{alm} are
\beeq
a_{lm}= \frac{2\pi}{a^4\bar{\rho}_\gamma }\int d^2\hat n 
~Y^*_{lm}(\hat n)\int dq ~q^3f(q,\hat n)\, .
\label{eqn:app:coefficients}
\eneq
The monopole coefficient is, therefore, related to the photon
density fluctuation~$\delta_\gamma$ as
\beeq
a_{00}=\sqrt{4\pi}\times {1\over4}\delta_\gamma^{\rm sync}=\sqrt{4\pi} \,\Theta^{\rm sync}~,
\eneq
and using eq.~\eqref{eqn:a_lm} the monopole transfer function is obtained as
\beeq
\TT_0(k)=\TTT_{\Theta}^{\rm sync}(k)~,
\eneq
where $\delta\rho_\gamma=\bar\rho_\gamma\delta_\gamma$ 
is related to the distribution function~$f(q,\hat n)$ as in eq.~\eqref{bgg}.
For the monopole transfer function, we need to use the gauge-invariant
expression~$\hat\Theta(\hat n)$, rather than the gauge-dependent 
expression~$\Theta(\hat n)$. In the comoving-synchronous gauge, where the
coordinate lapse vanishes, both values are identical.

The dipole coefficient is
\beeq
a_{10}=\sqrt{3\pi\over4}{2\over a^4\bar\rho_\gamma}\int d^3q~q\cos\theta
f(q,\hat n)=\sqrt{3\pi\over4}{s_z\over\bar\rho_\gamma}~,
\eneq
where the spatial energy flux of the photon distribution is defined as
\beeq
s^i:={2\over a^4}\int d^3q~q^if(q,\hat n)~.
\eneq
Using the fluid description for the photon energy-momentum tensor,
the spatial energy flux of the photon distribution can be expressed as
\beeq
s^i=(\bar\rho+\bar p)_\gamma u^i_\gamma~,
\eneq
in terms of the photon velocity~$u^i_\gamma$ 
measured in the observer rest frame. The photon velocity 
\beeq
u^i_\gamma=[e_i]^{\rm obs}_\mu u^\mu_\gamma=(v_{\rm obs}-v_\gamma)^{,i}~,
\eneq
is literally the relative velocity between the observer and the photon fluid
at the linear order, and it is evidently gauge-invariant, 
where $[e_i]^\mu_{\rm obs}$ is a spatial directional vector (or a spatial
tetrad) of the observer (see, e.g., 
\cite{Yoo_Grimm_Mitsou_Amara_Refregier_2018,Mitsou_2020}).
Finally, by using eq.~\eqref{eqn:a_lm}, the dipole transfer function is 
obtained as
\beeq
\TT_1(k)={k\over3}\left(\TTT_{v_\gamma}-\TTT_{v_{\rm obs}}\right)(k)~.
\eneq
In the comoving-synchronous gauge,
where the observer velocity vanishes, the dipole transfer function becomes
\beeq
\TT_1(k)={k\over3}\TTT_{v_\gamma}^{\rm sync}(k)~.
\eneq

The monopole and the dipole transfer functions~$\TT_0(k)$ and~$\TT_1(k)$
are obtained by using the above expressions and numerically evaluating
$\TTT_{\Theta}^{\rm sync}(k)$ and $\TTT_{v_\gamma}^{\rm sync}(k)$
at present day from the Boltzmann codes \textsc{class} and~\textsc{camb}.

\bibliography{paper.bbl}

\end{document}